\documentclass{aa}  
\usepackage{graphicx,color,epsfig}
\usepackage{times,natbib}
\usepackage{supertabular,multirow,longtable,rotating}
\usepackage[varg]{txfonts}

\def\kms{km\,s$^{-1}$}
\def\Ebv{\ensuremath{\mathrm{E}_\mathrm{(B-V)}}}

\def\ie{{\emph i.e.}}

\bibpunct{(}{)}{;}{a}{}{,}

\begin{document}

\title{VLT/X-shooter survey of near-infrared diffuse interstellar bands%
\thanks{Based on observations obtained with the VLT (ESO Programme 385.C-0720).}}  

\author{N.\,L.\,J. Cox\inst{1} \and J. Cami\inst{2,3} \and L. Kaper\inst{4} \and P. Ehrenfreund\inst{5}
  \and B.\,H. Foing\inst{6} \and B.\,B. Ochsendorf\inst{7} \and S.\,H.\,M. van Hooff\inst{4} \and F.~Salama\inst{8} }

\institute{ Instituut voor Sterrenkunde, KU\,Leuven, Celestijnenlaan 200D, Bus 2401, B-3001 Leuven, Belgium \\
\email{nick.cox@ster.kuleuven.be} 
\and Department of Physics and Astronomy, The University of Western Ontario, London, ON N6A 3K7, Canada 
\and SETI Institute, 189 Bernardo Ave, Suite 100, Mountain View, CA 94043, USA 
\and Sterrenkundig Instituut Anton Pannekoek, Universiteit van Amsterdam, Science Park 904, NL-1098, Amsterdam, The Netherlands 
\and Space Policy Institute, 1957 E Street, 20052 Washington DC, USA 
\and ESTEC, European Space Agency, Keplerlaan 1, NL-2201~AZ~~Noordwijk ZH, The Netherlands 
\and Leiden Observatory, Leiden University, Niels Bohrweg 2, NL-2333~CA~~Leiden, The Netherlands 
\and NASA AMES Research Centre, CA, United States}

\date{Received November 15, 2013, Accepted July 2, 2014}
 
\abstract{%
The unknown identity of the diffuse interstellar band (DIB) carriers poses one of the longest standing unresolved problems in
astrophysics. While the presence, properties, and behaviour of hundreds of optical DIBs between 4000~\AA\ and 9000~\AA\ have
been well established, information on DIBs in both the ultra-violet and near-infrared (NIR) ranges is limited.  
}{%
In this paper, we present a spectral survey of the NIR range, from 0.9~$\mu$m to 2.5~$\mu$m. Our observations were designed to detect new DIBs,
confirm previously proposed NIR DIBs, and characterise their behaviour with respect to known line-of-sight properties
(including the optical DIBs present in our spectra).  
}{%
Using the X-shooter instrument mounted on the ESO Very Large Telescope (VLT) we obtained medium-resolution spectra 
of eight known DIB targets and one telluric reference star, from 3000~\AA\ to 25\,000~\AA\ in one shot.  
}{%
In addition to the known 9577, 9632, 10780, 11797, and 13175~\AA\ NIR DIBs, we confirm 9 out of the 13 NIR DIBs that were
presented by Geballe and co-workers in 2011. 
Furthermore, we report 11 new NIR DIB candidates.
The strengths of the strongest NIR DIBs show a general correlation with reddening, \Ebv, but with a large scatter. 
Several NIR DIBs are more strongly correlated with the 5780~\AA\ DIB  strength than with \Ebv; 
this is especially the case for the 15268~\AA~DIB. 
The NIR DIBs are strong: the summed equivalent widths of the five strongest NIR DIBs represent a small percent 
of the total equivalent width of the entire average DIB spectrum (per unit reddening). 
The NIR DIBs towards the translucent cloud \object{HD 147889} are all weak with respect to the general trend.  
No direct match was found between observed NIR DIBs and laboratory matrix-isolation spectroscopic data of 
polycyclic aromatic hydrocarbons (PAHs).
}{%
The strong correlation between the 5780-15268 DIB pair implies that (N$f$)$_\mathrm{5780}$ / (N$f$)$_\mathrm{15268}$ = 14. 
However, the reduced strength of the 15268~\AA\ DIB in \object{HD 147889} rules out a common carrier for these two DIBs. 
Since the ionisation fraction for small PAHs in this translucent cloud is known to be low compared to diffuse clouds, 
the weakness of the 15268~\AA\ DIB suggests that an ionised species could be the carrier of this NIR DIB.
}
 

\titlerunning{Near-infrared diffuse interstellar bands}
\authorrunning{N.\,L.\,J.~Cox et al.}

\maketitle

\begin{table*}[th!]
\centering
\caption{Observed targets.}\label{tb:targets}
\begin{tabular}{lllllll}\hline\hline
Target 	     	& Spectral     & B-V	 & (B-V)$_0$\tablefootmark{b}	&\Ebv\tablefootmark{c}  & $A_V$      & $v_\mathrm{shift}$\tablefootmark{d} \\
		& Type\tablefootmark{a}& (mag) & (mag) 			& (mag)		   & (mag)  	     & (\kms)     \\  \hline
HD\,167785      &  B2\,V       & -0.09	 & -0.21$^1$,  -0.24$^2$ 	& 0.14  	   & 0.42 (0.05)     & -7.0       \\  
HD\,326306   	&  B1\,V       & -0.11	 & -0.23$^1$,  -0.26$^2$	& 0.14 	           & 0.42 (0.05)     &  8.0       \\   
HD\,153294   	&  B7\,Ib/II   & +0.03	 & -0.04$^1$,  -0.12$^2$	& 0.07, 0.15	   & 0.22 -- 0.47    & 14.0       \\   
HD\,152246   	&  O9\,Ib      & +0.12	 & -0.27$^1$,  -0.28$^2$	& 0.39 (0.03)      & 1.22 (0.03)     &  6.5       \\   
HD\,161056   	&  B1.5\,V     & +0.38	 & -0.22$^1$,  -0.25$^2$	& 0.62 (0.05)      & 1.90 (0.05)     & -3.5       \\   
HD\,161061   	&  O$^+$       & +0.62	 & -0.30$^1$,  -0.29$^2$	& 0.92 (0.03)      & 2.83 (0.04)     &  7.5       \\   
HD\,147889   	&  B2\,III/IV  & +0.71	 & -0.19$^1$,  -0.24$^2$	& 0.90 (0.06)      &  3.7 (0.1)      & 10.0       \\   
HD\,183143   	&  B7\,Iae     & +1.00	 & -0.04$^1$,  -0.04$^2$ 	& 1.04 (0.03)      & 3.97 (0.05)     & -5.0       \\   
4U\,1907$+$09  	&  O9\,Ia      & +3.2	 & -0.28$^1$,  -0.28$^2$	& 3.48 (0.03)      & 10.9 (0.05)     & 25.0       \\ \hline
\end{tabular}
\tablefoot{
\tablefoottext{a}{Spectral types taken from the Simbad database, except for 4U\,1907+09 taken from \citet{2005A&A...436..661C} and HD\,161061 taken from \citet{2011A&A...527A..34M}}.
\tablefoottext{b}{Intrinsic colours from (1) \citet{1994MNRAS.270..229W} and (2) \mbox{\citet{1970A&A.....4..234F}}.}
\tablefoottext{c}{\Ebv\ values are derived from $B$ and $V$ magnitudes (Simbad) and intrinsic colours, (B-V)$_0$.
We took the mean of the intrinsic colours from (1) and (2) with uncertainties including inaccuracy, by one step,
in Morgan-Keenan spectral-luminosity classification. We adopt \Ebv = 0.10~mag for HD\,153294.$A_V = R_V \times~\Ebv$. 
We adopt $R_V = 3.1$ except for HD\,147889 ($R_V = 4.1$) and HD\,183143 ($R_V = 3.8$). Uncertainties in parentheses.}
\tablefoottext{d}{The heliocentric radial velocity of the strongest components in the atomic line of 
\ion{K}{i} ($\lambda_\mathrm{rest}$ = 7698.974~\AA) 
used to shift the spectra to the respective ``LSR'' rest wavelengths. Corrections were verified for consistency 
using additional interstellar lines of the \ion{Na}{i}\,D doublet, CH ($\lambda_\mathrm{rest}$ = 4300.303~\AA), and CH$^+$ ($\lambda_\mathrm{rest}$ = 4232.548~\AA).}
}
\end{table*}

\nocite{2011A&A...527A..34M}
\nocite{2005A&A...436..661C}

\section{Introduction}

\begin{figure*}[htp!]
\centering
\resizebox{\hsize}{!}{
 \includegraphics[width=0.33\textwidth]{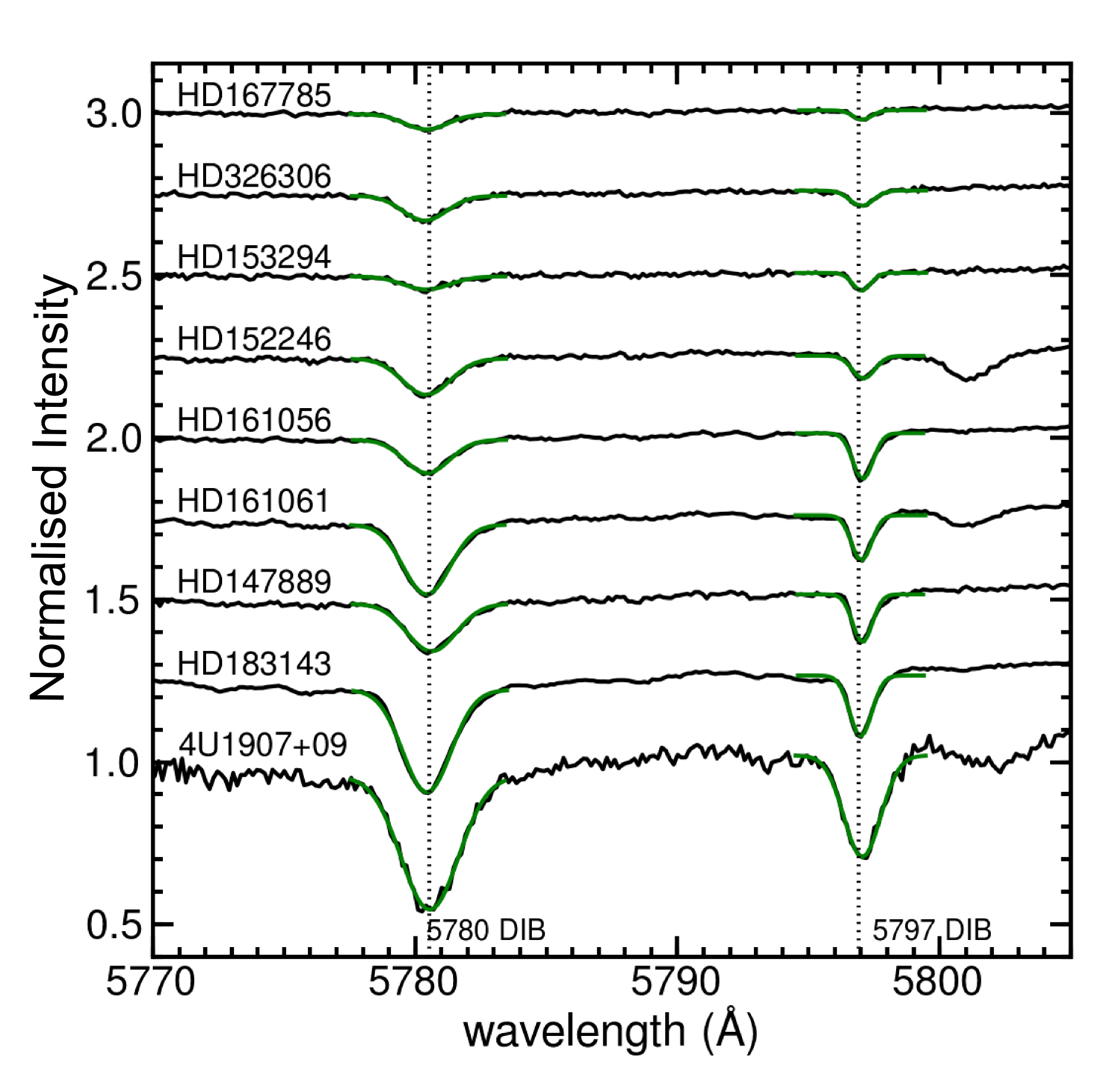}
 \includegraphics[width=0.33\textwidth]{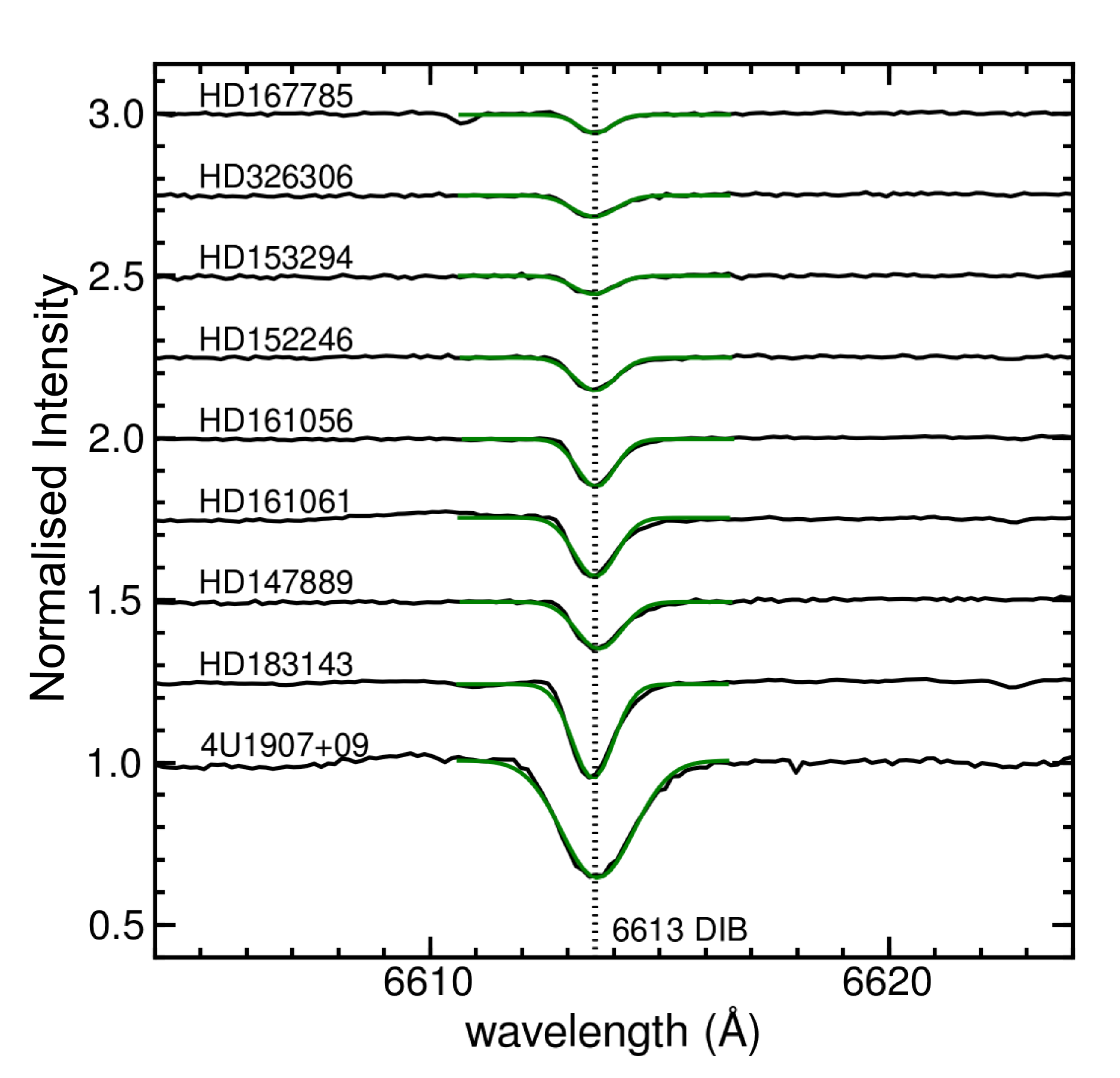}
 \includegraphics[width=0.33\textwidth]{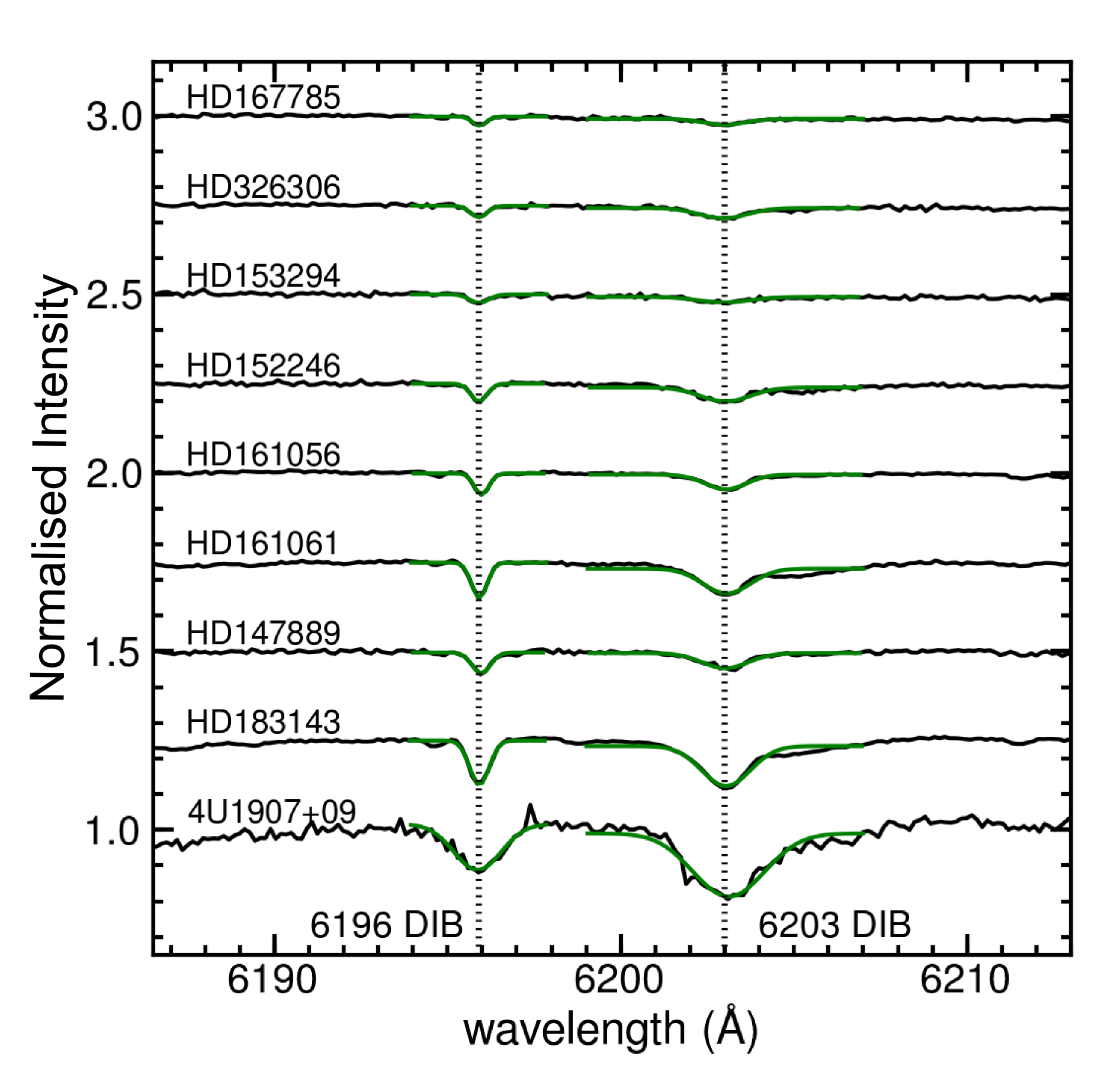}}
 \caption{The strong DIBs at 5780~\AA\ and 5797~\AA\ (left), 6613~\AA\ (middle), and 6196 \& 6203~\AA\ (right) 
 are shown for the observed targets. Solid green curves represent the Gaussian model fit to the data.}
 \label{fig:opticalDIBs}
\end{figure*}

Diffuse interstellar bands (DIBs) are observed throughout the Universe, and represent a major, albeit unidentified, fraction of
interstellar matter in the Milky Way and beyond (e.g. \citealt{2002SnowReview}, \citealt{2014IAUS..297...41C}, and
references therein).  Several hundred strong and weak DIBs have been reported between 4000~\AA\ and 9000~\AA\
(\citealt{1994A&AS..106...39J}; \citealt{2000A&AS..142..225T}; \citealt{2009ApJ...705...32H}), while only a few are known in the NIR
longwards of 9000~\AA\ (\citealt{1990Natur.346..729J}; \citealt{1994Natur.369..296F}; \citealt{2011Natur.479..200G}). 
Despite intensive study - and debate - of a number of propositions and reported laboratory matches, not a single DIB carrier has been
unambiguously identified.  For a review of the various proposed candidate carriers see the seminal work by
\citet{1995ARA&A..33...19H}; \citet{2006JMoSp.238....1S} provides a detailed discussion of more recent developments in the 
quest for the DIB carriers.

One important question arising from DIB studies is whether DIB features are also present at shorter ultra-violet (UV) and longer IR
wavelengths. The presence/absence of DIBs in the UV and IR is important for testing the PAH-DIB hypothesis
(\citealt{1996ApJ...458..621S}, \citealt{1995A&A...299..213E}, \citealt{2011EAS....46..349C}), because charged polycyclic aromatic 
hydrocarbons (PAHs) containing more than about 20 carbon atoms are known to have their main electronic transitions in the NIR range, 
whereas neutral PAHs and small charged PAHs absorb predominantly at ultra-violet and visible wavelengths, respectively 
(see \citealt{2008ARA&A..46..289T}). \citet{2011ApJ...728..154S} and \cite{2011A&A...530A..26G} searched for the main transition 
of neutral PAHs in astronomical optical spectra, but no matches were found and only upper limits could be set to their abundance. 
On the other hand, PAHs containing more than 40 carbon atoms are quite stable, could be fully hydrogenated 
(\citealt{2000A&A...363L...5V}), and have strong transitions in the (near-)infrared. 
Hence, NIR spectroscopic (DIB) surveys may reveal their presence in the interstellar medium. Several ongoing and future large NIR
spectroscopic surveys include NIR DIBs and hence an accurate catalogue of these features, including an understanding of their 
behaviour with respect to other interstellar features, is warranted. 
For example, \citet{2014IAUS..297...68Z,2014arXiv1406.1195Z} present first results on two NIR DIBs in SDSS-III/APOGEE spectra 
and their Galactic distribution.
Notwithstanding this potential importance, the NIR wavelength range (0.9~$\mu$m--2.5~$\mu$m) has so far received little attention. 
\citet{1994MNRAS.268..705A} have searched, without success, for the 13175~\AA\ band in dense environments.  
\citet{2011Natur.479..200G} presented the discovery of a dozen NIR DIB candidates longwards of 1.5~$\mu$m.
These new finding have revived interest for this spectral range in context of the properties and chemical complexity 
of the ISM in general, and DIB carriers in particular.
Searches for DIBs in the near-UV at wavelengths shortwards of 3300~\AA\ have thus far shown that there are apparently far fewer DIBs, 
if any, in the near-UV than in the optical (\citealt{1997AcA....47..225G,2001ApJ...555..472W}; Bhatt et al. in preparation).

This paper presents the results of an exploratory VLT/X-shooter survey of NIR DIBs in the spectra of a sample of reddened early-type
stars. This works seeks to confirm and extend the results on the presence, properties and behaviour of NIR DIBs in diffuse/translucent 
interstellar clouds. The observations and data reduction are briefly discussed in Sect.~\ref{sec:obs}. 
Then, in Sect.~\ref{sec:results} we present spectra, equivalent width and full-width-at-half-maximum (FWHM) measurements of 
known visible and NIR DIBs; we also identify additional NIR DIBs. The NIR DIB properties in terms of strength
and width are discussed in Sect.~\ref{sec:nirdib-properties} together with correlations of DIB strengths with dust (reddening) 
and with other DIBs. In Sect.~\ref{sec:laboratory} the observations are compared with previous laboratory measurements. 
Section~\ref{sec:hd147889} discusses the measured DIB properties for the translucent cloud towards HD\,147889 in context of 
the expected PAH population. The main conclusions of this survey are summarised in Sect.~\ref{sec:conclusion}.

\section{Optical and NIR spectroscopy with VLT/X-shooter}\label{sec:obs}

We selected a small sample of bright reddened early-type stars whose lines of sight give rise to a range in magnitudes of extinction. 
The diffuse clouds in these sightlines are also exposed to the interstellar radiation field in varying degrees.  
This optimises the ability to detect new NIR DIBs and study their behaviour in the ISM. 
The observed targets are listed in Table~\ref{tb:targets} with their interstellar reddening, \Ebv, visual extinction, $A_V$, 
and spectral type. To obtain a spectroscopically unbiased view of NIR DIBs, spectra were obtained with X-shooter, a wide-band 
intermediate-resolution spectrograph at the ESO {\it Very Large Telescope} (\citealt{2006SPIE.6269E..98D};
\citealt{2011A&A...536A.105V}). X-shooter has three separate arms (optical light paths) - \emph{UVB}, \emph{VIS}, and 
\emph{NIR} - each with their own optimised optical and system design. 
In this way it is possible to obtain in one shot a nearly contiguous spectrum from 3000~\AA\ to 24800~\AA.  
The spectra were reduced, separately for each arm, using the \textsl{esorex} X-shooter pipeline (\citealt{2010SPIE.7737E..56M}). 
Standard parameters were adopted for preparing the calibration files. The final science spectra were reduced using non-default 
stricter cosmic ray hit rejection criteria which significantly improved the final spectra by removing spurious glitches. 
In this paper we focus on the NIR spectra which cover the range from 9940~\AA\ to 24790~\AA. 
A 0.4\arcsec\ slit, with a fixed length of 11\arcsec, was used in the ``Auto Nod on Slit Mode'' yielding a spectral resolving 
power of $\sim$10\,000 in the NIR arm. This corresponds to an instrumental FWHM, measured from telluric lines, of
$\sim$1.3~\AA\ in the range 1.1-1.3~$\mu$m, $\sim$1.7~\AA\ in the range 1.5-1.7~$\mu$m, and $\sim$1.9~\AA\ at 
1.8~$\mu$m\footnote{All quoted rest wavelengths are those measured in air unless otherwise specified. 
DIB rest wavelengths given by \citet{2011Natur.479..200G} are in vacuum.}. 
The telluric correction tool Spextool (see \citealt{2003PASP..115..389V} and \citealt{2004PASP..116..362C}) was
used to remove atmospheric absorption lines from the observed NIR spectra by way of dividing the spectra with a telluric standard
spectrum. In order to have an independent verification of the procedure and the final corrected spectrum we use both a computed
atmospheric model and the spectrum of an unreddened B star. 
The hydrogen lines are removed from the telluric standard as part of the Spextool telluric correction procedure. 
The telluric correction is prone to introduce significant residuals in regions with dense forests of strong telluric absorption lines. 
The following regions are most severely affected: 13500~\AA\ -- 14800~\AA, 17900~\AA\ -- 19000~\AA, 19950~\AA\ -- 20250~\AA, 
20500~\AA\ -- 20750~\AA\, 23150~\AA\ -- 24800~\AA. However, even in those cases the narrow residuals can sometimes be 
distinguished from intrinsic stellar and interstellar lines.

\begin{figure}[t!]
\centering
 \includegraphics[width=1.0\columnwidth]{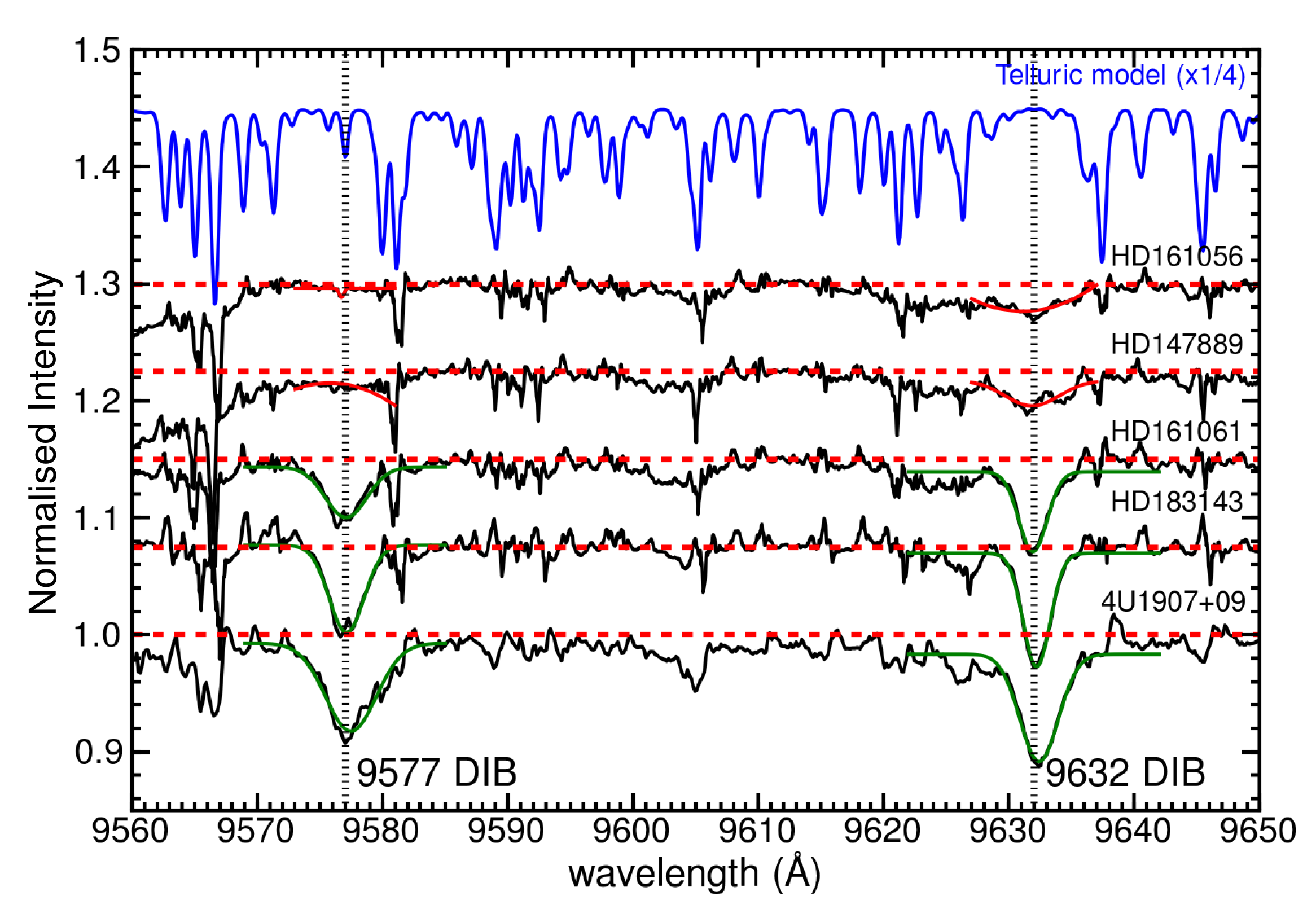}
 \caption{NIR DIBs at 9577~\AA\ and 9632~\AA\ (assigned to the fullerene cation C$_{60}^+$ by 
 \citealt{1994Natur.369..296F}) are shown for the 
 reddened X-shooter sample of sightlines.
 The spectra have been corrected for the radial velocity of the ISM (including heliocentric correction).
 The vertical dotted lines indicate the rest wavelength for the respective DIBs. Dashed horizontal lines 
 correspond to the normalised continua. Spectra are vertically offset for clarity.
 The atmospheric model spectrum (blue) is shown at the top (normalised and vertically compressed) in order to appreciate
 and evaluate the telluric correction.
 Solid green curves show the Gaussian model fits.
 The weak feature at 9633~\AA\ towards HD\,161056 and HD\,147889 indicates the presence of a weak stellar \ion{Mg}{ii} line.
 This has been noted before by \citet{1997A&A...317L..59F} and estimated by \citet{2000MNRAS.317..750G} to have an 
 equivalent width of 50~m\AA\ for spectral type B2\,III and 80~m\AA\ for B7\,I.
 These authors assigned two additional features at 9603~\AA\ and 9625~\AA, present also in our spectra, to stellar \ion{He}{i}.}
 \label{fig:C60}
\end{figure}

\begin{figure}[t!]
\centering
 \includegraphics[width=.9\columnwidth]{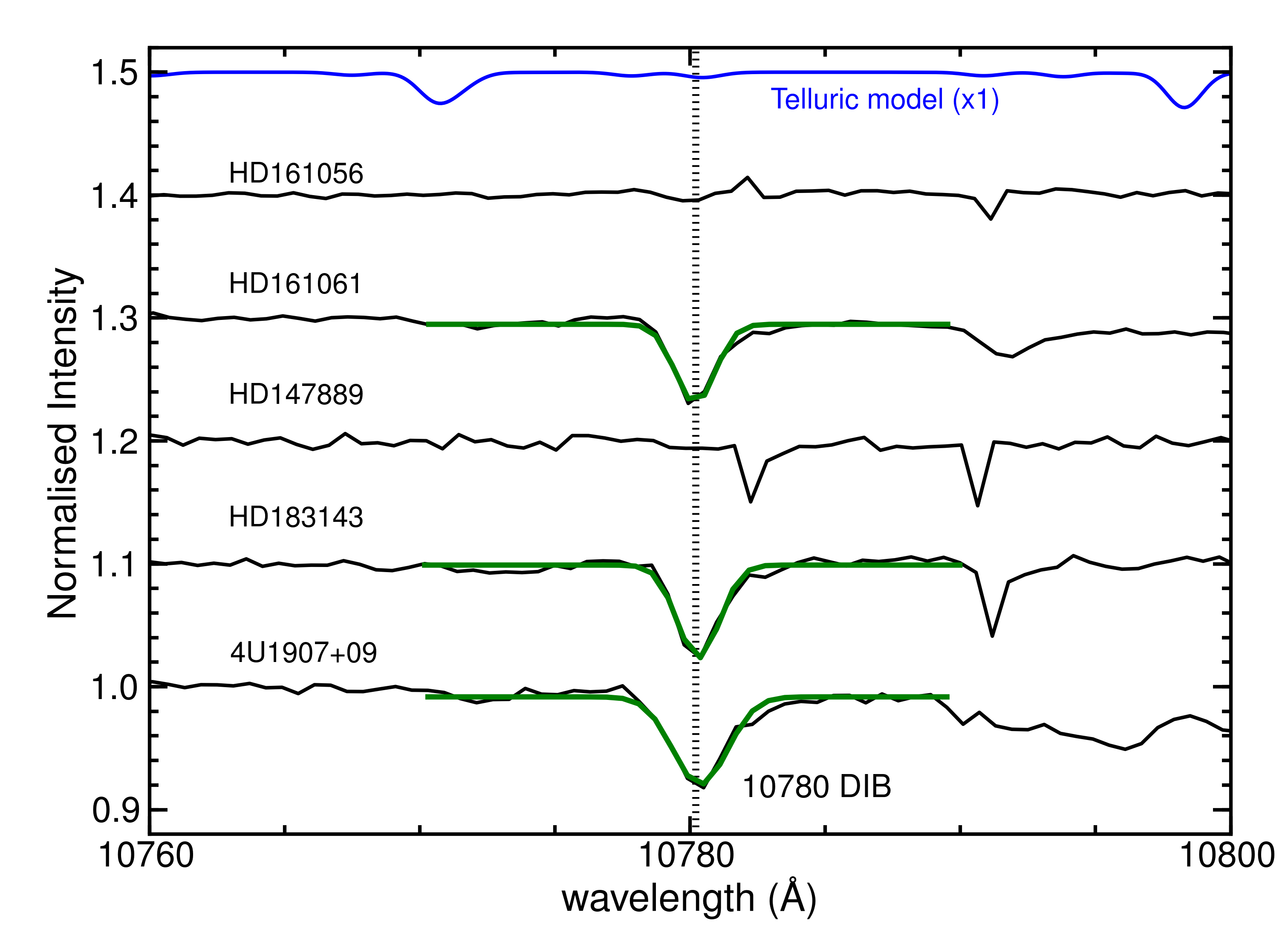}
 \caption{NIR DIBs at 10780~\AA\ \citep{2007A&A...465..993G} are shown for the reddened X-shooter sample of sightlines.
 The spectra have been corrected for the presence of telluric lines and subsequently for the radial velocity of the ISM.
 The dashed lines indicate the rest wavelength for the respective DIB. 
 Solid green curves show the Gaussian model fits and the blue solid curve displays the telluric model spectrum.}
 \label{fig:10780}
\end{figure}

\section{Diffuse interstellar bands}\label{sec:results}

\subsection{Measurements of optical diffuse interstellar bands}\label{subsec:opticalDIBs}

Before studying the NIR DIBs, we measured the equivalent widths of several strong DIBs in the optical range 
(X-shooter \emph{VIS} arm; see Fig.~\ref{fig:opticalDIBs} and Table~\ref{tb:nirdib_measurements}). 
The average equivalent width per unit reddening for the seven strong DIBs at 5780, 5797, 5849, 6196, 6203, 6270, 
and 6613~\AA\ is 1.6~\AA/\Ebv.

We also used \ion{Na}{i} and \ion{K}{i} absorption lines to determine the radial structure of the ISM, providing the opportunity
to shift the observed optical and NIR DIBs to a common restframe. 
Once applied, this correction ensures that wavelength positions of the interstellar lines coincide between different
lines of sight, while telluric and stellar lines display a shift due to the radial velocity with respect to the ISM.

\subsection{Detection of NIR diffuse interstellar bands}\label{sec:NIRDIBs}

\begin{figure*}[thp!]
\centering
\resizebox{\hsize}{!}{%
 \includegraphics{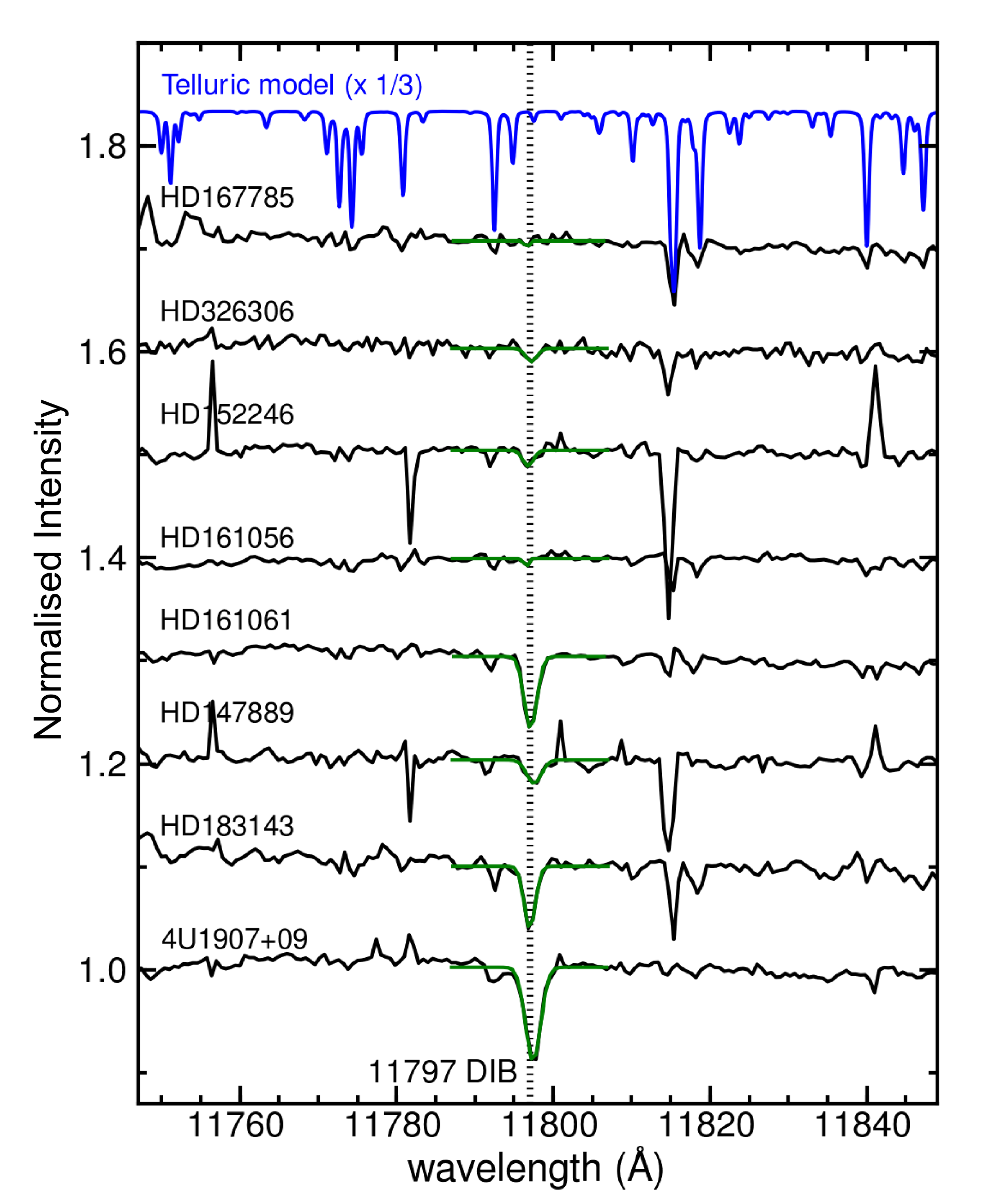}
 \includegraphics{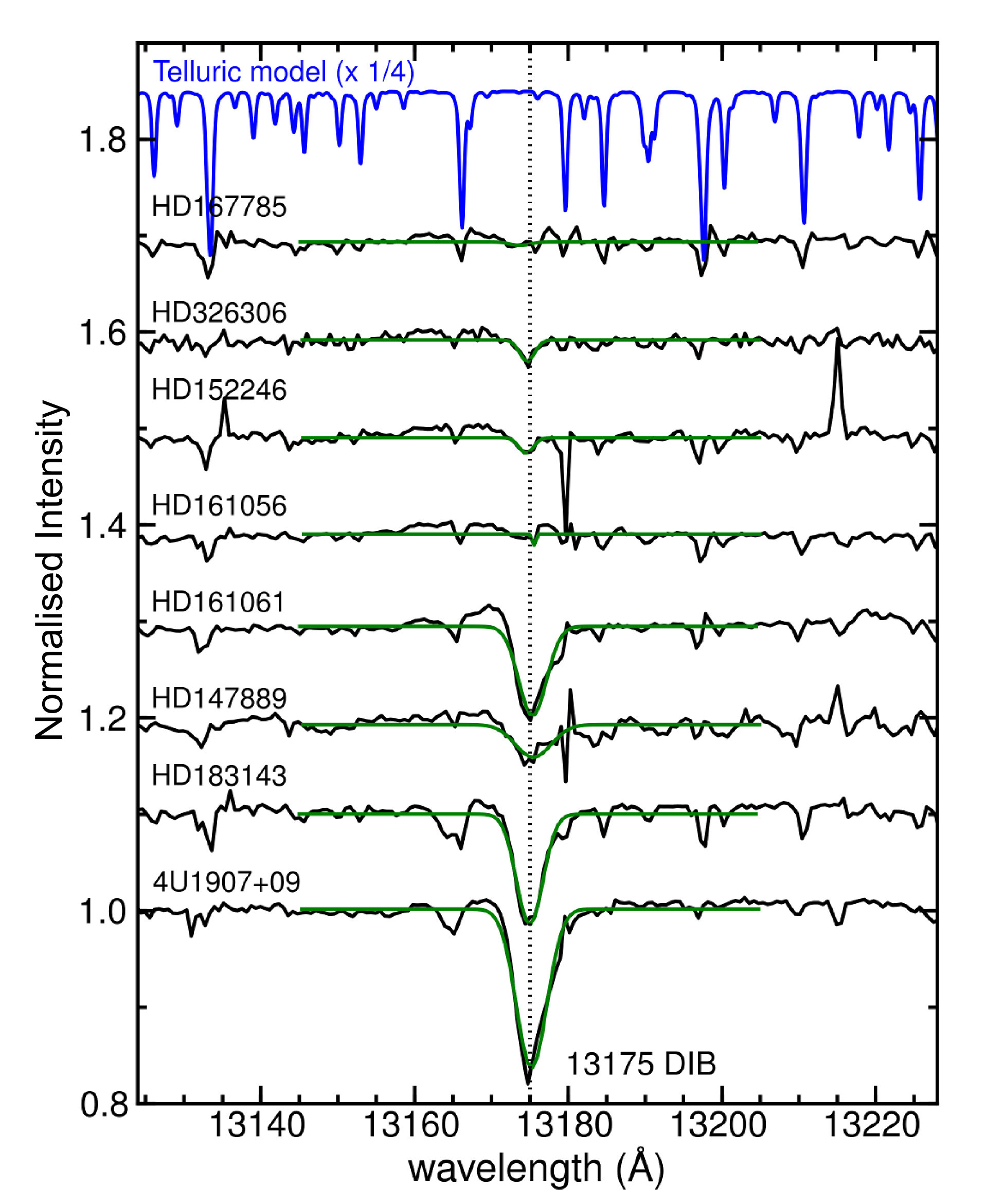}
 \includegraphics{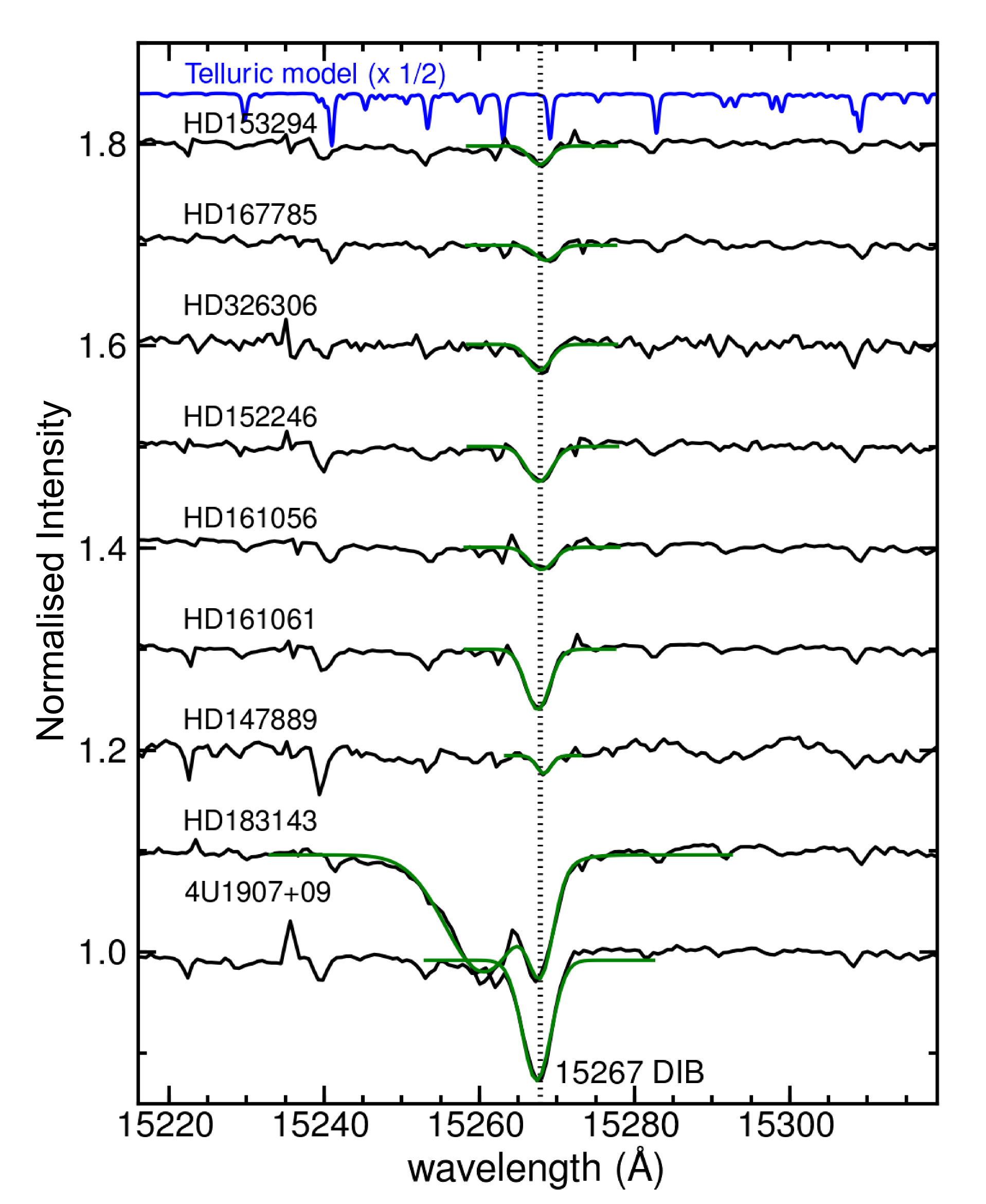}
}
 \caption{NIR DIBs at 11797~\AA\ (left panel), 13175~\AA\ \citep[middle panel;][]{1990Natur.346..729J}, and
   15268~\AA\ \citep[right panel;][]{2011Natur.479..200G} are shown for the X-shooter sample of sightlines.
 The spectra have been corrected for the presence of telluric lines and subsequently for the radial velocity of the ISM.
 The dashed lines indicate the rest wavelength for the respective DIBs. 
 Solid green curves show the Gaussian model fits and the blue solid curve displays the telluric model spectrum.
}
\label{fig:JoblinGeballe}
\end{figure*}

\begin{figure*}[tph!]
\centering
\resizebox{\hsize}{!}{%
\includegraphics{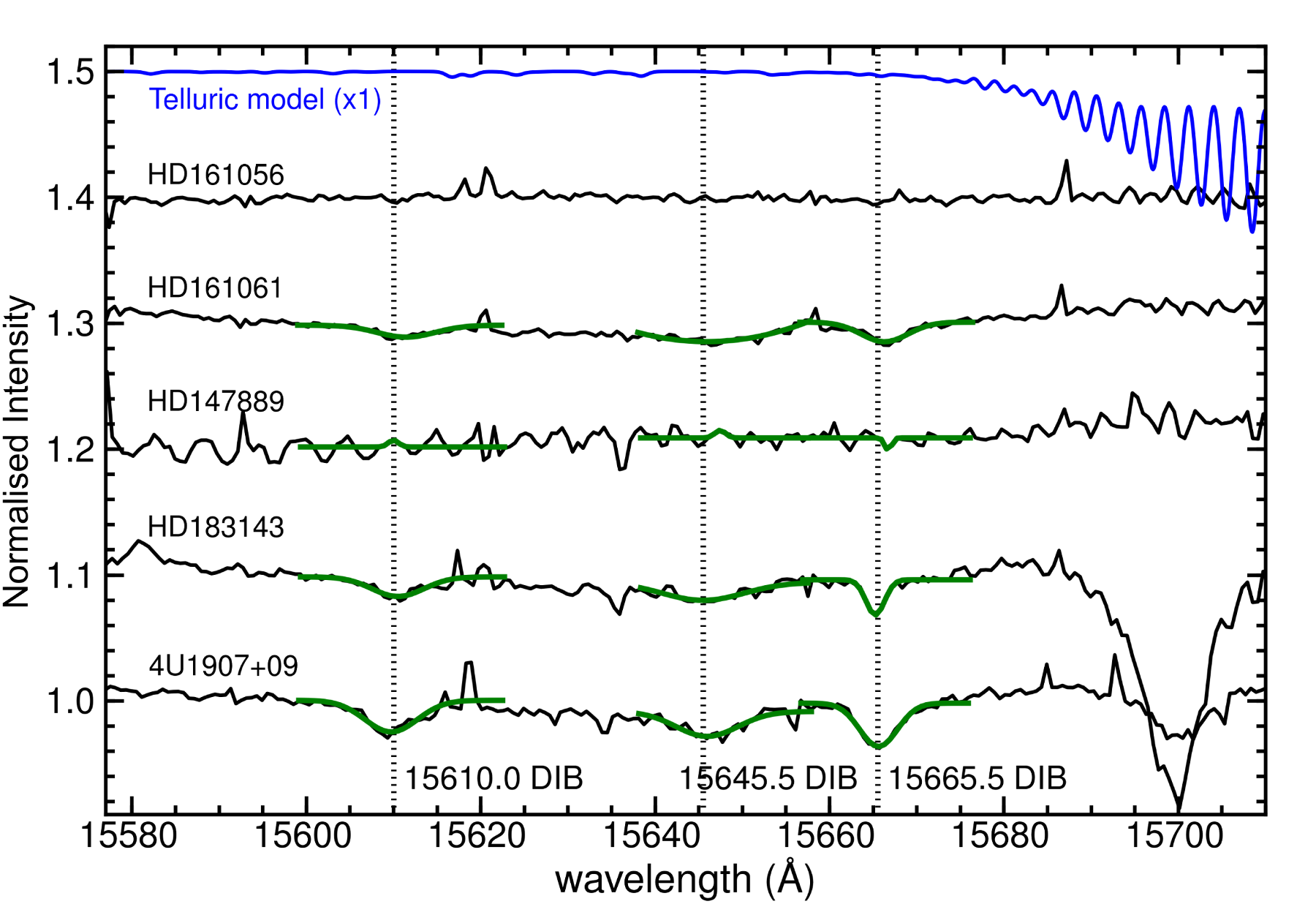}
\includegraphics{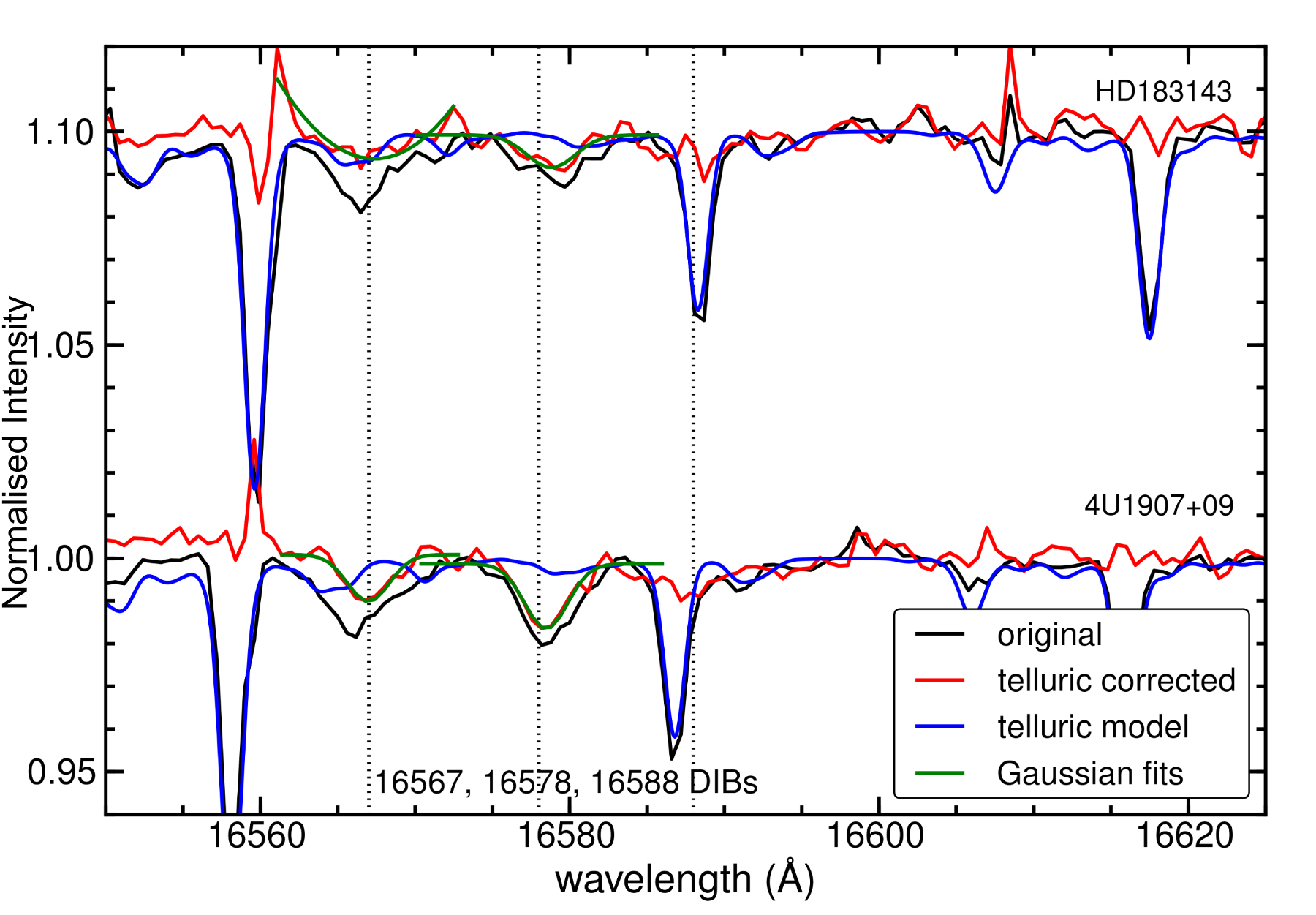}}
 \caption{Same as Fig.~\ref{fig:JoblinGeballe} but for the 15610, 15646, and 15666~\AA\ NIR DIBs 
 towards five reddened targets (left panel) and for the 16567, 16578, and 16588~\AA\ NIR DIBs 
 in two reddened lines of sight (right panel). 
 Solid green curves show the Gaussian model fits and the blue solid curve displays the telluric model spectrum.
 In the right panel the original (normalised) spectra are shown in black while the telluric corrected spectrum 
 is overplotted in red.}
 \label{fig:GeballeDIBs}
\end{figure*}

\citet{1990Natur.346..729J} reported the first two NIR DIBs at 11797~\AA\ and 13175~\AA\ towards several reddened Galactic lines of sight. 
Next, the NIR DIBs at 9577~\AA\ and 9632~\AA\ were detected in a larger sample of bright OB stars (\citealt{1994Natur.369..296F,2000MNRAS.317..750G}).
As part of their spectral atlas of OB stars \citet{2007A&A...465..993G} reported two unidentified features at 10780 and 10792~\AA\
in 9 lines of sight. The absorption band at 10780~\AA\ was found to correlate well with reddening (cf. their Fig.~4).
Recently, \citet{2011Natur.479..200G} reported the detection of 13 new NIR DIBs, measured towards the Galactic Centre, 
4 of which are also present in the average spectrum of the heavily reddened Cygnus OB2 sightlines. 
Figs.~\ref{fig:C60} to~\ref{fig:GeballeDIBs} exhibit most of these NIR DIBs in our sample of low- to moderately-high reddening lines of sight. 
The spectra have been corrected for heliocentric motion as well as for the radial velocity of the main interstellar
cloud in the line-of-sight according to the optical atomic absorption lines of \ion{Na}{i} and \ion{K}{i} 
(Sect.~\ref{subsec:opticalDIBs}).

Owing to the limited signal-to-noise ratio and intrinsic weakness most of the known NIR DIBs are only detected in the three 
lines of sight with a visual extinction above 2~mag (\ie\ HD\,161061, HD\,183143, 4U\,1907+09). 
The strong and narrow $\lambda\lambda$11797, 13175, 15268~DIBs are detected in some other sightlines as well. 
Of the thirteen new NIR DIBs reported in \citet{2011Natur.479..200G} we detect three in three or more sightlines. 
Three others are detected in at least two sightlines while another three are detected only towards 4U\,1907. 
Four NIR DIBs reported by Geballe and co-workers could not be confirmed (see below). 
The presence and absence of NIR DIBs in this X-shooter survey is summarised as follows:

\begin{itemize}

\item The NIR DIBs at 9577 and 9632~\AA\ \citep{1994Natur.369..296F} are detected in three lines of sight (Fig.~\ref{fig:C60}).

\item The NIR DIB at 10780~\AA\ \citep{2007A&A...465..993G} is detected in three lines of sight (Fig.~\ref{fig:10780}).

\item The NIR DIBs at 11797 and 13175~\AA\ (\citealt{1990Natur.346..729J}) are detected in seven lines of sight (Fig.~\ref{fig:JoblinGeballe}).

\item The strong NIR DIB at $\lambda_\mathrm{rest}$\,15267.8~\AA\ reported by \citet{2011Natur.479..200G} is detected in nine
      sightlines (Fig.~\ref{fig:JoblinGeballe}).

\item The bands at $\lambda_\mathrm{rest}$\,15610 and 15665~\AA\ \citep{2011Natur.479..200G} are detected in 
      three lines of sight (Fig.~\ref{fig:GeballeDIBs}).

\item The $\lambda_\mathrm{rest}$\,15646 and 16588~\AA\ DIBs \citep{2011Natur.479..200G} are seen 
      in two sightlines (Fig.~\ref{fig:GeballeDIBs}).

\item The blended NIR DIB complex at 16560-16590~\AA\ reported by \citet{2011Natur.479..200G} is resolved into three separate components.

\item The NIR DIBs at $\lambda_\mathrm{rest}\,16226.7$, 16567.0, 16588.0, and 17802.7~\AA\ 
      reported by \citet{2011Natur.479..200G} are confirmed only in the spectrum of 4U\,1907+09.

\item The shallow/broad bands at 15225, 15990, 17758, and 17930~\AA\ reported by \citet{2011Natur.479..200G} are not detected. 
      The 15990, 17758, and 17930~\AA\ DIBs are situated in regions with significant telluric line residuals in the X-shooter 
      spectra presented here. Particularly, the 15225, 15990, and 17758~\AA\ DIBs are among the weakest reported by 
      \citet{2011Natur.479..200G}, and scaled to the extinction of $\sim$10 visual magnitudes towards 4U\,1907+09, would have 
      depths of only 1.5, 0.5, and 1.5~\%, respectively (to be compared to the weak, but narrow, tentative NIR DIBs detected 
      at a level of 2.0\%\ in this line-of-sight).

\end{itemize}

\begin{figure*}[pht!]
\centering
\resizebox{.75\hsize}{!}{\includegraphics{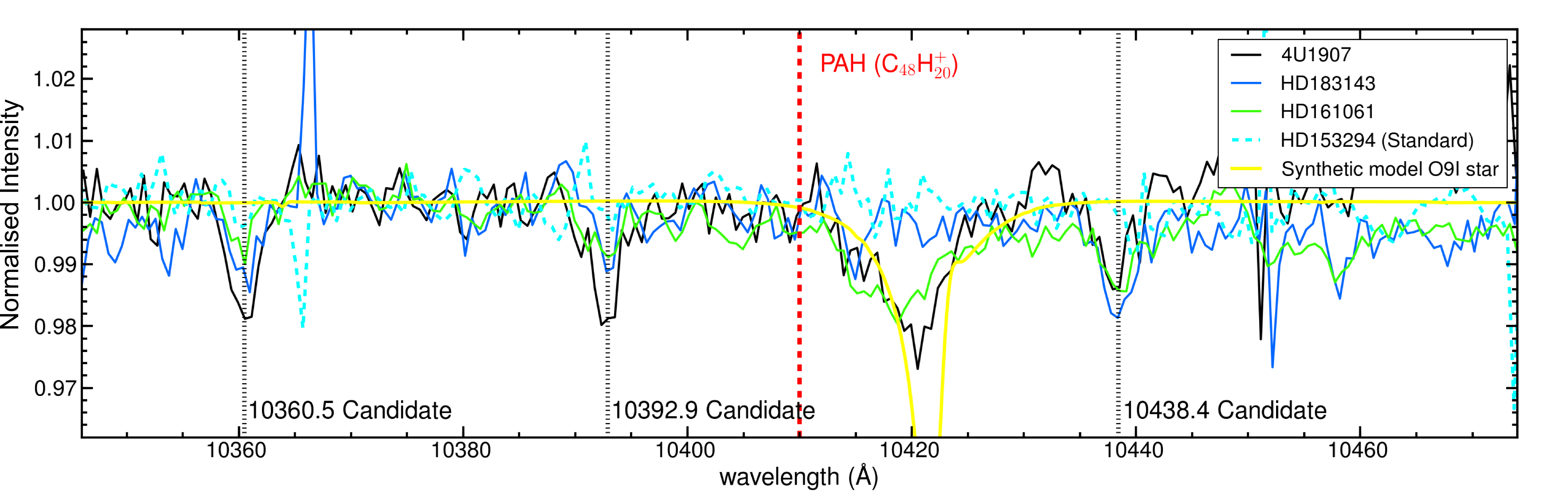}}
\resizebox{.75\hsize}{!}{\includegraphics{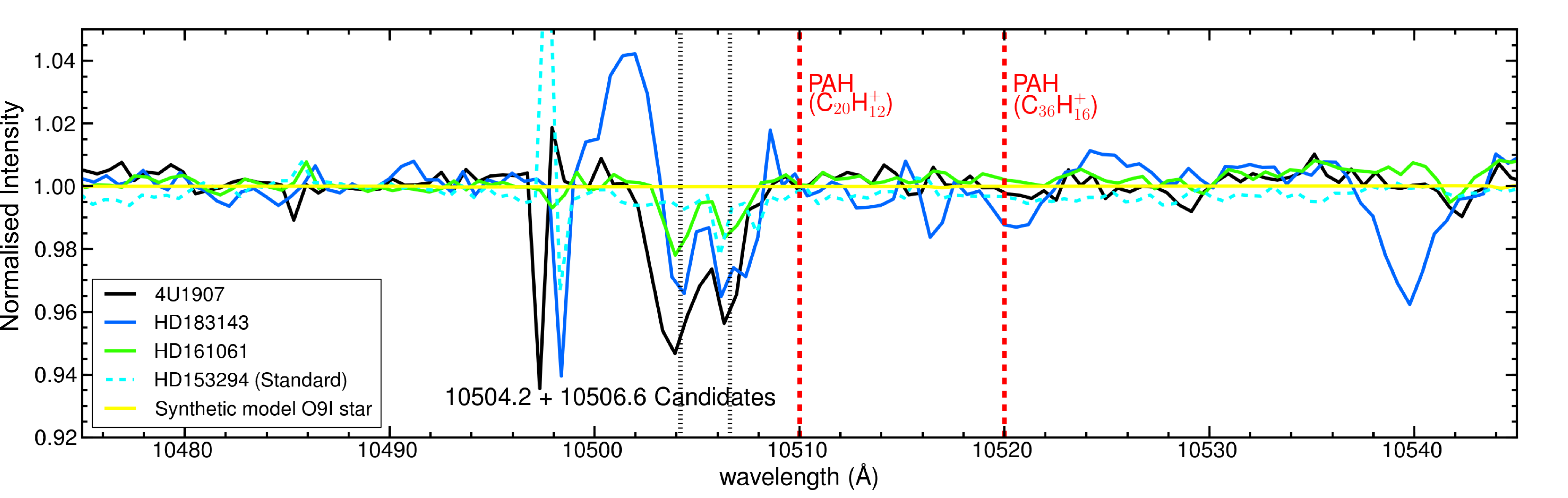}}
\resizebox{.75\hsize}{!}{\includegraphics{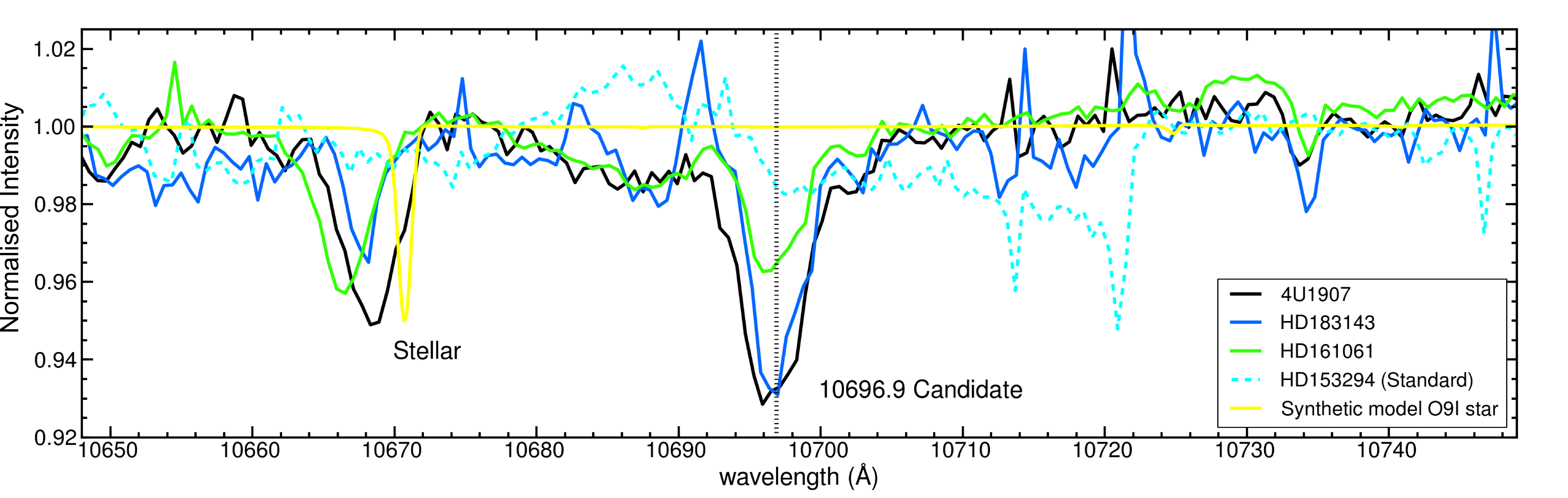}}
\resizebox{.75\hsize}{!}{\includegraphics{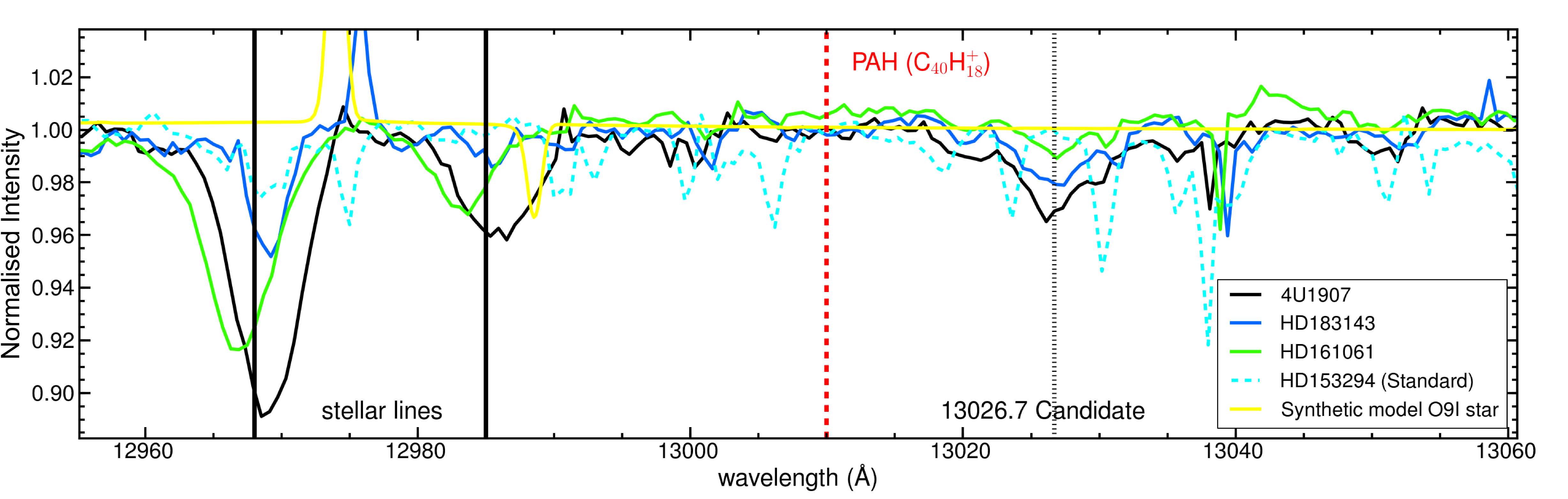}}
 \caption{Candidate NIR DIBs at $\lambda\lambda$10360, 10392, 10438,
   10503, 10506, 10696, 
   and 13026~\AA\ for reddened sightlines towards HD\,161061 (green), HD\,183143 (blue), 
   and 4U\,1907+09 (black). The spectra have been shifted to the interstellar rest frame.
   The dotted black lines indicate the rest wavelength for the candidate NIR DIBs.  
   Solid black lines in the bottom panel indicate the position of stellar lines. 
   The dashed red lines correspond to the laboratory positions of main absorption bands of several
   mid-sized PAHs reported by \citet{2005ApJ...629.1188M}. 
   The peaks of the stellar lines (at 10422, 10670, 10830, 12969, and 12984~\AA) do not coincide as a 
   result of differences in radial velocity of the stars. 
   The telluric standard spectrum of HD\,153294 (dashed cyan) and stellar atmosphere model of an O9Ia star
   (solid yellow) are overplotted for reference.}
\label{fig:candidates} 
\end{figure*}

\begin{figure*}[pht!]
\centering
\resizebox{.75\hsize}{!}{\includegraphics{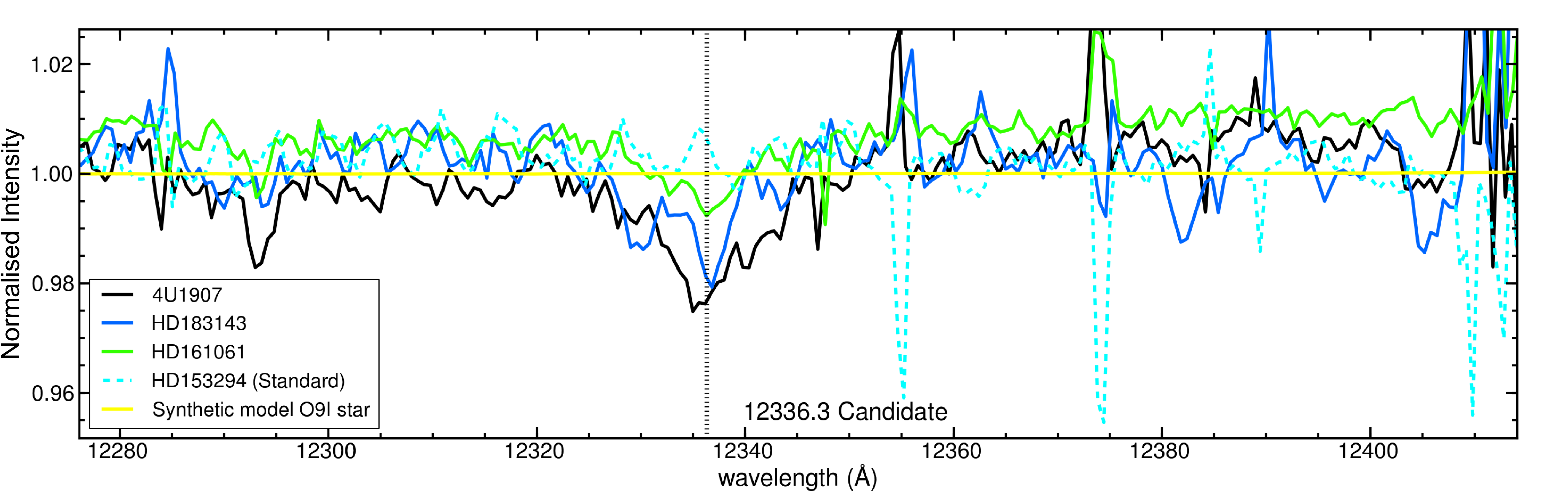}}
\resizebox{.75\hsize}{!}{\includegraphics{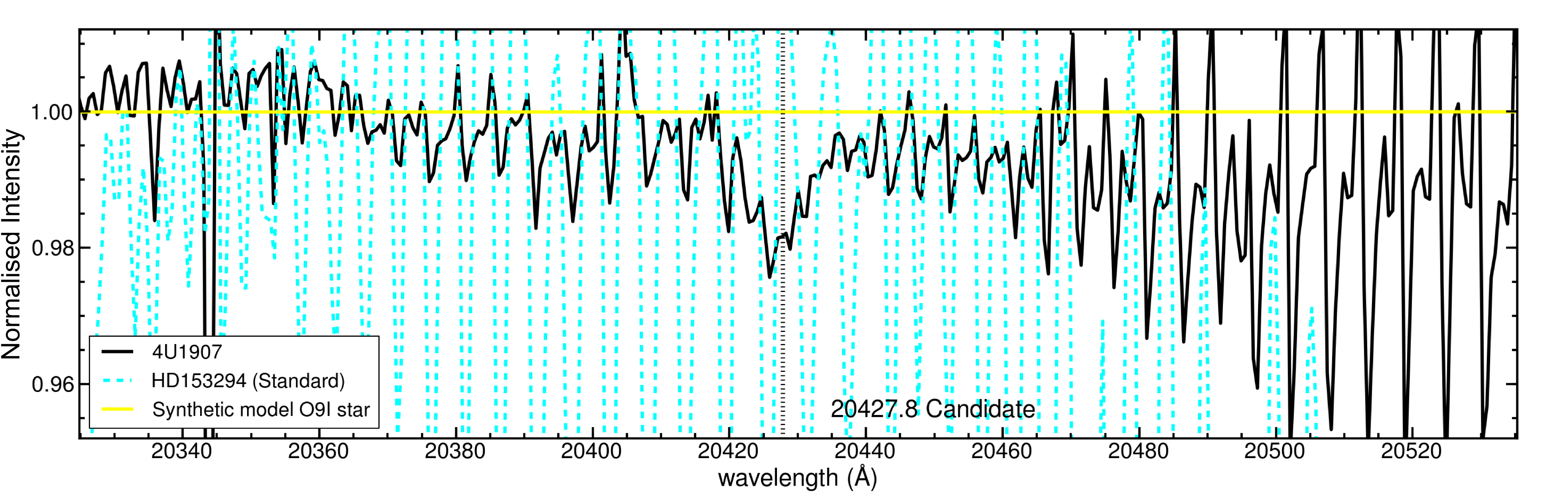}}
\resizebox{.75\hsize}{!}{\includegraphics{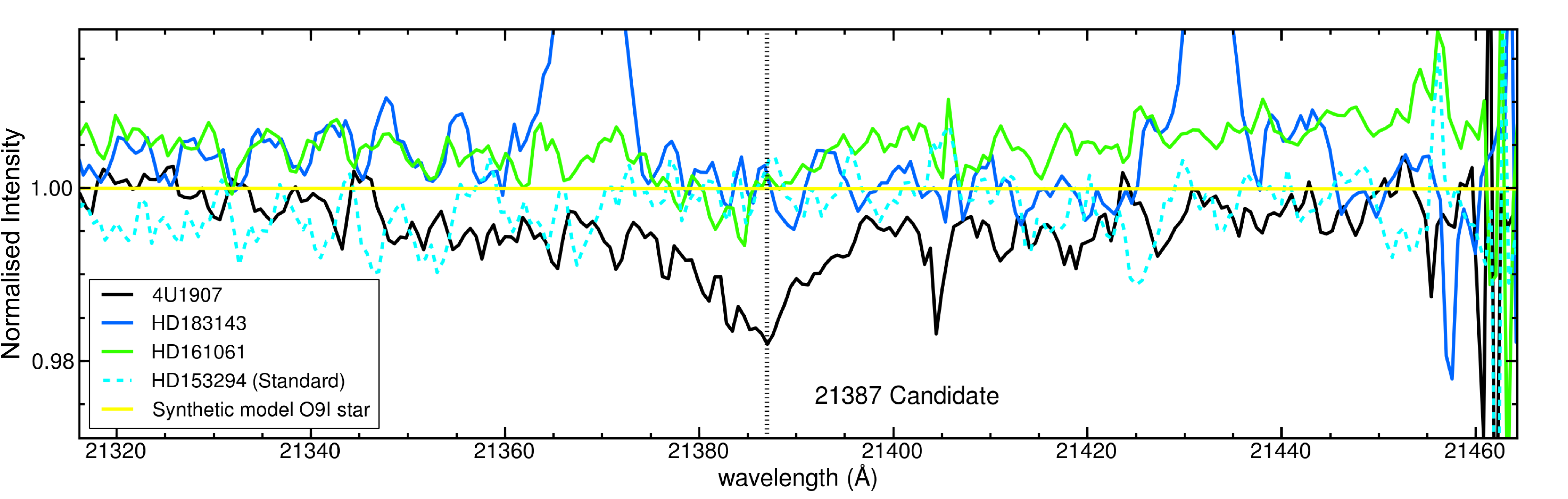}}
\resizebox{.75\hsize}{!}{\includegraphics{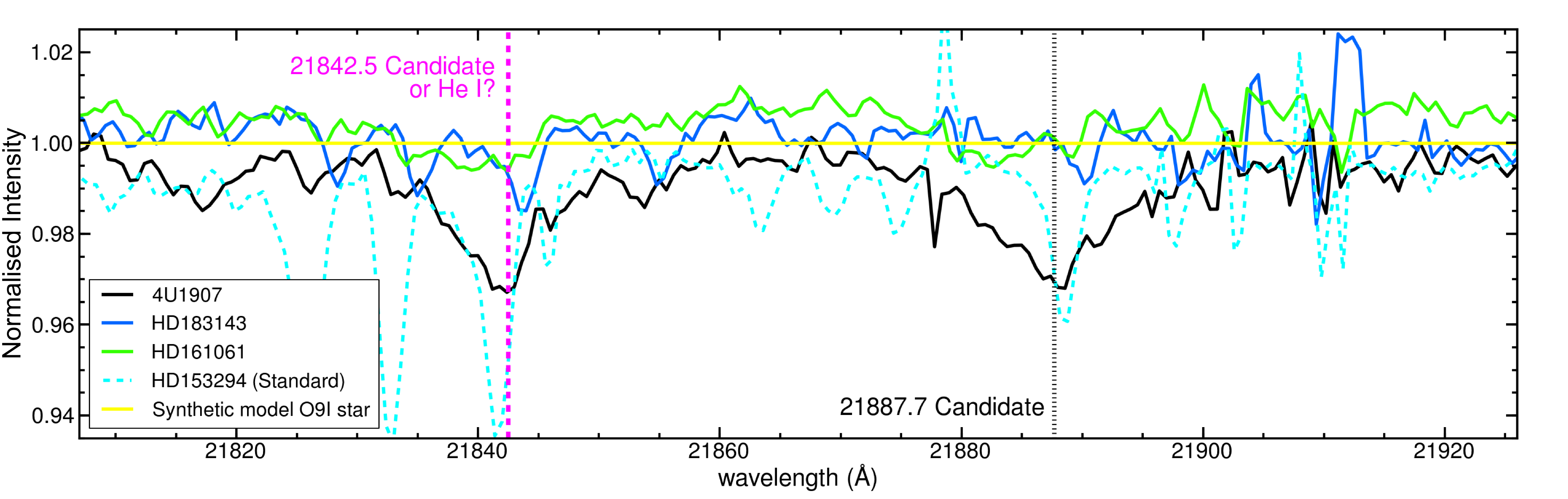}}
 \caption{Tentative NIR DIB candidates at 12336.3~\AA, 20427.8~\AA, 21387.0~\AA, 21842.5~\AA, and 21887.7~\AA\ (from top to bottom)
   towards the reddened star 4U\,1907 (and HD\,183143 and HD\,161061 where applicable).  
   Dotted vertical lines indicate the position of the tentative NIR DIBs. 
   The telluric standard spectrum of HD\,153294 (dashed cyan) and stellar atmosphere model of an O9Ia star (solid yellow) 
   are overplotted for reference. The feature at 21842.5~\AA\ is possibly (a blend with) a stellar \ion{He}{i} line.}
 \label{fig:tentative_candidates}
\end{figure*}

In addition to these previously reported NIR DIBs we report seven new NIR DIB candidates in the spectra of the three most 
reddened stars (\ie\ HD\,161061, HD\,18314, 4U\,1907+09). Similar to the known NIR DIBs these new candidates are not 
detected towards HD\,147889 (Sect.~\ref{sec:hd147889}). Candidates are listed in Table~\ref{tb:nirdib_measurements} (bottom section)
and respective spectra are shown in Fig.~\ref{fig:candidates}.  These seven DIB candidates have similar strengths per unit 
reddening as the known NIR DIBs and their rest wavelengths coincide in different sightlines. 
Even though the latter criterion provides a strong argument in favour of an interstellar origin, we remain cautious in view of
possible limiting factors such as ($i$) insufficient signal-to-noise ratio for spectra with low reddening, \Ebv~$<$~1.0~mag, 
($ii$) imperfect removal of telluric lines, ($iii$) confusion with stellar lines, and ($iv$) the small number of available 
sightlines with moderate/high-resolution NIR spectra of appropriate interstellar sightlines. 
Additional possible interstellar features at 12336~\AA, 20428~\AA, 21387~\AA, and 21843~\AA\ are present in the spectrum of
4U\,1907+09, but not, or only marginally, in spectra of HD\,183143 and HD\,161061. 
These lines are also not seen in the spectrum of the non-reddened standard star, so that a stellar origin can be excluded.
For completeness these tentative interstellar features are shown in Fig.~\ref{fig:tentative_candidates}.  
Further studies at high continuum sensitivity for a large sample are needed to provide further constraints on the presence of 
DIBs in the NIR range from 1~$\mu$m to 3~$\mu$m, although in practice results will be limited by how accurate both telluric
and stellar atmosphere lines can be accounted for.

\nocite{2005ApJ...629.1188M}

\begin{table*}[hpt!]
\centering
\caption{Equivalent widths in m\AA\ (uncertainties in parentheses) of NIR DIBs in reddened sightlines observed with X-shooter.}
\label{tb:nirdib_measurements}
\begin{tabular}{lp{1.4cm}p{1.4cm}p{1.4cm}p{1.4cm}p{1.4cm}p{1.4cm}p{1.4cm}p{1.4cm}p{1.4cm}p{0.0cm}}\hline\hline
$\lambda_\mathrm{DIB}$\tablefootmark{a} 
			& HD\,167785	& HD\,326306	& HD\,153294	& HD\,152246	& HD\,161056	& HD\,161061	& HD\,147889	& HD\,183143	& 4U\,1907	&       \\ 
\hline \\
5780			& 98 (4)	& 161 (6)	& 90 (7)	& 253 (8)	& 220 (5)	& 486 (9)	& 349 (8)	& 742 (12)	& 1172 (31)	& 	\\
5797			& 26 (2)	& 53 (3)	& 48 (2)	& 75 (3)	& 135 (3)	& 133 (4)	& 137 (3)	& 197 (6)	& 540 (16)	& 	\\
5849			& 5 (3)		& 19 (2)	& 22 (2)	& 21 (1)	& 59 (1)	& 50 (2)	& 74 (2)	& 61 (2)	& 186 (10)	& 	\\
6196			& 13 (1)	& 19 (1)	& 17 (2)	& 30 (1)	& 32 (1)	& 56 (2)	& 40 (2)	& 87 (2)	& 218 (21)	& 	\\
6203			& 26 (3)	& 51 (3)	& 30 (4)	& 73 (6)	& 62 (2)	& 122 (8)	& 71 (3)	& 202 (9)	& 465 (26)	& 	\\
6270			& 17 (3)	& 37 (2)	& 14 (2)	& 50 (3)	& 52 (2)	& 100 (4)	& 54 (6)	& 170 (5)	& 405 (12)	&	\\
6613			& 50 (3)	& 79 (2)	& 62 (2)	& 115 (2)	& 152 (2)	& 207 (5)	& 177 (4)	& 329 (5)	& 703 (13)	&	\\ 
\\\hline\\
9577     		& --		& --		& --		& --		& --		& 180 (5)	& $\leq$90	& 260 (5)	& 380 (6)	&\\
9632   			& --		& --		& --		& --		& --		& 195 (4)	& $\leq$120	& 263 (3)	& 360 (5)	&\\
10780                   & --		& --		& --		& --		& --		& 119 (6)   	& --		& 148 (8)   	& 181 (11)   	&\\
11797   		& --		& 21 (7)	& --		& 27 (8)	& --		& 145 (7)	& 56 (17)	& 110 (9)	& 224 (12)	&\\
13175    		& --		& 53 (12)	& --		& 40 (24)	& --		& 419 (33)	& 203 (53)	& 493 (36)	& 819 (36)	&\\
15268   		& 54 (14)	& 93 (15)	& 60 (16)	& 146 (18)	& 85 (19)	& 231 (17)	&$\approx$40 (7)& 326 (28)	& 554 (38)	&\\
15610   		& --		& --		& --		& --		& --		& 79 (30)	& -- 		& 116 (43)	& 189 (66)	&\\
15646   		& --		& --		& --		& --		& --		& --		& -- 		& 153 (88)	& 174 (57)	&\\
15666   		& --		& --		& --		& --		& --		& --		& -- 		& 78 (10)	& 188 (15)	&\\
16227			& --		& --		& --		& --		& --		& --		& -- 		& --		& 696 (148)	&\\
16567   		& --		& --		& --		& --		& --		& --		& -- 		& --		& 37 (6)	&\\
16578   		& --		& --		& --		& --		& --		& --		& -- 		& 30 (10)	& 53 (8)	&\\
16588   		& --		& --		& --		& --		& --		& --		& -- 		& --    	& 18 (5)	&\\
17803			& --		& --		& --		& --		& --		& --		& -- 		& --		& 502 (166)	&\\
\\\hline\\
10361   		& --		& --		& --		& --		& --		& 18 (4)	& --		& 46 (32)	& 69 (10)	&\\		
10393   		& --		& --		& --		& --		& --		& 25 (8)	& --		& 19 (5)	& 56 (9)	&\\
10438   		& --		& --		& --		& --		& --		& 28 (4)	& --		& 56 (9)	& 65 (15)	&\\
10504\tablefootmark{e,f}& --		& --		& --		& --		& --		& 36		& --		& 58		& 136		&\\
10507\tablefootmark{e,f}& --		& --		& --		& --		& --		& 29		& --		& 261		& 68		&\\
			& --		& --		& --		& --		& --		& (66)\tablefootmark{d}		& (295)\tablefootmark{d} & (203)\tablefootmark{d} &\\
10697\tablefootmark{f}  & --		& --		& --		& --		& --		& 127 (19)  	& --		& 262 (36)   	& 329 (22)   	&\\
13027\tablefootmark{e}  & --		& --		& --		& --		& --		& 71 (15)  	& --		& 171 (37)   	& 222 (41)   	&\\
\\
\hline
\end{tabular}
\tablefoot{
\tablefoottext{a}{Measured rest peak wavelengths in air are 
9577.1$\pm$0.2, 9632.0$\pm$0.2, 10780.3$\pm$0.2, 11797.0$\pm$0.5, 13175.0$\pm$0.5, 
15267.8$\pm$0.4, 15610.0$\pm$0.5, 15645.5$\pm$0.5, 15665.5$\pm$0.5, 16226.7$\pm$0.5, 16567.0$\pm$0.5, 16578.0$\pm$0.5, 
16588.0$\pm$0.5, 17802.7$\pm$0.5, 
10360.5$\pm$0.3, 10392.9$\pm$0.3, 10438.4$\pm$0.1, 10504.2$\pm$0.2, 10506.6$\pm$0.2, 10696.9$\pm$0.2, 13026.7$\pm$0.5.}
\tablefoottext{d}{Integrated equivalent width for blended ``W-shape'' features at 10504~\AA\ \& 10506~\AA. 
Absorption lines for HD\,183143 are contaminated by a strong emission feature.}
\tablefoottext{e}{Complex profile; possible blended lines.}
\tablefoottext{f}{No sensible uncertainties could be determined for individual components.}
}
\end{table*}

\subsubsection{Measurements of NIR DIBs}

We fitted a Gaussian to the NIR DIBs and determined the equivalent width, FWHM and rest-frame wavelength from the best-fit parameters. 
Equivalent widths were also measured by numerical integration, yielding similar results within the uncertainties. 
The results for the detected NIR DIBs in each sightline are given in Table~\ref{tb:nirdib_measurements}.

\section{NIR DIB properties}\label{sec:nirdib-properties}

Given the number of NIR DIB detections, we can now compare some of their properties in more 
detail to those of the optical DIBs. We will in particular look at their widths, strengths and correlations.

\subsection{Rest wavelengths and DIB profiles}

The moderate resolving power ($R~\sim~10\,000$) of X-shooter is insufficient to resolve smaller substructure 
or weak asymmetries in the NIR DIB absorption line profiles. However, compared to the work by 
\citet{2011Natur.479..200G}, our lines of sight suffer much less from line broadening due to blending of 
multiple velocity components in the line of sight that are Doppler shifted as a result of Galactic rotation.
With the exception of 4U\,1907, Doppler broadening is not expected to exceed 10~\kms~(or 0.5~\AA\ at 15\,000~\AA ); 
thus, the observed FWHM from 1.9~\AA\ to 7.0~\AA\ (see Table~\ref{tb:FWHM-NIRDIBs}) correspond to intrinsic 
FHWM\footnote{FWHM$_\mathrm{intrinsic}$ = $\sqrt{\mathrm{FHWM}_\mathrm{observed}^2 - \mathrm{FHWM}_\mathrm{instrumental}^2}$}
from 1.4~\AA\ (20~\kms) to 6.8~\AA\ (77~\kms).

Owing to this lack of Galactic rotation Doppler shift line broadening and the precise measurement of the line-of-sight radial 
velocity (see~\ref{subsec:opticalDIBs}), we can derive accurate rest wavelengths (reported in air) for the NIR DIBs listed in 
Table~\ref{tb:nirdib_measurements}. 
The significant Doppler broadening towards 4U\,1907 has been discussed in \citet{2005A&A...436..661C}. 

The measured widths (FWHM) of the NIR DIBs are listed in Table~\ref{tb:FWHM-NIRDIBs}. The NIR DIBs at 15268, 15665, 16567,
and 16578~\AA\ are narrow and have widths comparable to the earlier reported NIR DIBs at 9577, 9632, 11797, and 13175~\AA. 
We find that the widths of the 15268 and 15665~\AA\ DIBs are smaller than those reported by \citet{2011Natur.479..200G}, 
suggesting that Doppler velocity broadening affects their measurements. This is not the case for the broad 15610 and 17800~\AA\ 
NIR DIBs whose widths are similar to those by \citet{2011Natur.479..200G}. 
For the three reddened lines of sight with a narrow interstellar velocity structure we furthermore find that 
FWHM(11797) = 1.9 $\pm$ 0.4~\AA\ (somewhat narrower than the 2.7 $\pm$ 0.3~\AA\ reported by \citealt{1990Natur.346..729J}) 
and FWHM(13175) = 4.5 $\pm$ 1.2~\AA\ (comparable to the 4.0 $\pm$ 0.5~\AA\ measured by \citealt{1990Natur.346..729J}). 

Finally, it is interesting that the widths of the 11797 and 13175~\AA\ NIR DIBs in the single cloud line-of-sight towards
HD\,147889 are similar to those towards the multicloud object 4U\,1907. However, the NIR DIBs are weak in this sightline, 
and consequently uncertainties are large.

\subsection{DIB strengths}

Table~\ref{tb:nirdib_measurements} lists the equivalent widths for the NIR DIBs in the highly reddened lines of sight. 
It is clear that the detected NIR DIBs are fairly strong. The average NIR DIB strength per unit reddening, W(DIB)/\Ebv\ 
ranges from 0.07~\AA~mag$^{-1}$ to 0.35~\AA~mag$^{-1}$ in our lines of sight while the \emph{total} NIR DIB equivalent 
width per unit reddening (for the five strongest NIR DIBs) ranges from $\sim$0.6~\AA~mag$^{-1}$ to 1.4~\AA~mag$^{-1}$. 
This is similar to the total strength of the seven strongest optical DIBs, although the wavelength dependence in the 
equivalent width measurement favours the NIR DIBs equivalent width (in wavelength units) by a factor of two to three.

For an effective wavelength of 1.5~$\mu$m and assuming optically thin line absorption, we thus find that 
$f\times N_{\rm NIR DIB}$ = 3--7$\times$ 10$^{11}$~cm$^{-2}$ for all NIR DIBs combined; $f$ is then an effective oscillator strength. 
For the strongest optical DIBs ($\lambda_\mathrm{effective} = 6000~\AA$), $f\times N_\mathrm{DIB} = 5 \times 10^{12}$~cm$^{-2}$. 
Thus, if oscillator strengths are similar for both optical and NIR DIBs (as is e.g. the case for molecule no. 25 discussed in 
Sect.~\ref{sec:laboratory}), the abundance of NIR DIB carriers is an order of magnitude lower than that of the carriers of
the strongest optical DIBs.  If, however, the optical and NIR DIBs arise from the same species in the same state, their respective
column densities should be equal and hence the oscillator strengths of the NIR DIBs are an order of magnitude smaller. 
Note however that such a situation is not likely, given the imperfect correlation between NIR and optical DIBs (see below).

\subsection{DIB strength correlations}

In an effort to understand the relation between the NIR DIB carriers and those of the optical DIBs, we will look at
correlations between DIB strengths and reddening, and mutual correlations between the NIR DIBs and optical DIBs. This can only be
done in a meaningful way with a large enough sample of sightlines; we therefore only include the $\lambda\lambda$11797, 13175, and
15268 NIR DIBs for which we have at least six measurements; there are too few data points for the other NIR DIBs. 

\begin{table}[hpt!]
\centering
\caption{FWHM (in \AA; uncertainties in parentheses) of NIR DIBs in reddened sightlines observed with X-Shooter.}
\label{tb:FWHM-NIRDIBs}
\begin{tabular}{lp{1.4cm}p{1.4cm}p{1.4cm}p{1.4cm}}\hline\hline
$\lambda_\mathrm{DIB}$\tablefootmark{a} & \multicolumn{4}{c}{FWHM (\AA)} \\
     			& HD\,161061	& HD\,147889	& HD\,183143	& 4U\,1907	\\
\hline \\
9577             	& 3.9 (0.12)   	& --	  	& 3.3 (0.07)    & 4.8 (0.07)    \\
9632             	& 2.7 (0.05)   	& --     	& 2.5 (0.03)    & 3.6 (0.06)    \\
10780                   & 1.7 (0.1)    	& --		& 1.8 (0.1)     & 2.4 (0.2)     \\
11797            	& 1.9 (0.1)    	& 2.2 (0.7)    	& 1.7 (0.1)     & 2.3 (0.1)     \\
13175             	& 4.2 (0.3)    	& 5.5 (1.5)    	& 4.0 (0.3)     & 4.7 (0.2)     \\
15268            	& 3.6 (0.3)   	& 2.2 (0.4)     & 3.7 (0.3)     & 4.4 (0.3)     \\
15610            	& 7.8 (3.0)	& -- 		& 7.0 (2.6) 	& 7.1 (2.5) 	\\
15646            	& --		& -- 		& 11.0 (5.8)	& 8.3 (2.7) 	\\
15666            	& 6.3 (1.5)	& -- 		& 2.5 (0.3) 	& 5.1 (0.4) 	\\
16227			& --		& --		& --		& 33 (7)    	\\
16567            	& --		& -- 		& --        	& 3.2 (0.5) 	\\
16578            	& --		& -- 		& 3.7 (1.2) 	& 3.3 (0.5) 	\\
16588            	& --		& -- 		& --		& 2.9 (0.9) 	\\
17803			& --		& -- 		& --		& 13 (4)        \\
\\ \hline \\
10361          		& 2.0 (0.5)	& --		& 3.3 (2.3) & 3.1 (0.5)\\		
10393           	& 3.2 (1.0)     & --		& 1.4 (0.4) & 2.8 (0.5) \\
10438           	& 2.6 (0.4)   	& --		& 3.5 (0.5) & 4.0 (0.9) \\
10504\tablefootmark{b}	& 1.4     	& --		& 1.2       & 2.3 \\
10507\tablefootmark{b}	& 1.5     	& --		& 4.8       & 1.5 \\
10697          		& 3.8 (0.6)    	& --		& 3.9 (0.5) & 4.9 (0.3) \\
13027\tablefootmark{b}  & 4.7     	& --		& 5.5       & 5.7 \\
\hline
\end{tabular}
\tablefoot{
\tablefoottext{a}{Measured rest peak wavelengths in air are given in Table~\ref{tb:nirdib_measurements}.}
\tablefoottext{b}{Complex profile; possible blended lines or telluric residuals. 
No sensible uncertainties could be determined for (individual) components.}
}
\end{table}

\begin{table}[hpt!]
\centering
\caption{Equivalent widths (uncertainties in parentheses) for the 11797, 13175, and 15268~\AA\ NIR DIBs reported previously 
in the literature.}\label{tb:joblin}
\begin{tabular}{p{1.7cm}llll}\hline\hline
Target 	     		      & \Ebv	      & \multicolumn{3}{c}{Equivalent widths (m\AA)} 		\\  
	     		      & (mag)	      & 11797           & 13175 	  & 15268		\\ \hline
HD\,24398\tablefootmark{b}    & 0.34	      & 32 (9)        	& 50 (20)	  & --	        	\\ 
HD\,223385\tablefootmark{b}   & 0.59	      & 280 (38)       	& 460 (34)        & --	        	\\ 
HD\,20041\tablefootmark{b}    & 0.73	      & 34 (4)         	& 262 (35)     	  & --	        	\\ 
HD\,183143\tablefootmark{b}   & 1.04 	      & 169 (29)	& 451 (37)	  & -- 			\\ 
HD\,194279\tablefootmark{b}   & 1.22	      & 336 (54)      	& 560 (55)        & --	        	\\ 
BD+40\,4223\tablefootmark{a}  & 1.93 	      & -- 		& 860 (60)        & --	        	\\ 
BD+40\,4220\tablefootmark{b}  & 2.00	      & 394 (44)       	& 870 (60)        & --	        	\\ 
GCS3-2\tablefootmark{a}       & $\sim$6.5     & --		& --		  & 1500 (100)     	\\ 
qF362\tablefootmark{a}	      & 7.0 (0.6)     & --		& 3130 (100)      & --	        	\\ 
\hline
\end{tabular}
\tablefoot{
\tablefoottext{a}{Equivalent width measurements from \citet{2011Natur.479..200G}. 
Visual extinction for GCS3-2, qF362, and BD+40\,4223 is $\sim$20, 21.7$\pm$1.7, and 5.97~mag, respectively.}		    
\tablefoottext{b}{Equivalent width measurements from \citet{1990Natur.346..729J}.
HD\,183143 is the only X-shooter line-of-sight also observed by Joblin et al. 
The equivalent width of the 11797~\AA\ DIB observed with X-shooter is 40\% smaller, 
while those for the 13175~\AA\ DIB agree within the uncertainties.}
}
\end{table}

\subsubsection{Correlation with reddening}

The correlation coefficients\footnote{We always imply Pearson linear correlation coefficients in this work.} $r$ between the DIB
strengths and reddening range from 0.80 (for the 6203 and 6270~\AA\ DIBs) to 0.95 (for the 5797~\AA\ DIB) for the optical DIBs
in our sample confirming the well established positive but not perfect correlation between W and \Ebv\ 
(e.g.  \citealt{1993ApJ...407..142H}, \citealt{1997A&A...326..822C}, \citealt{2011ApJ...727...33F}).

\begin{figure*}[tph!]
\centering
\resizebox{0.45\hsize}{!}{\includegraphics{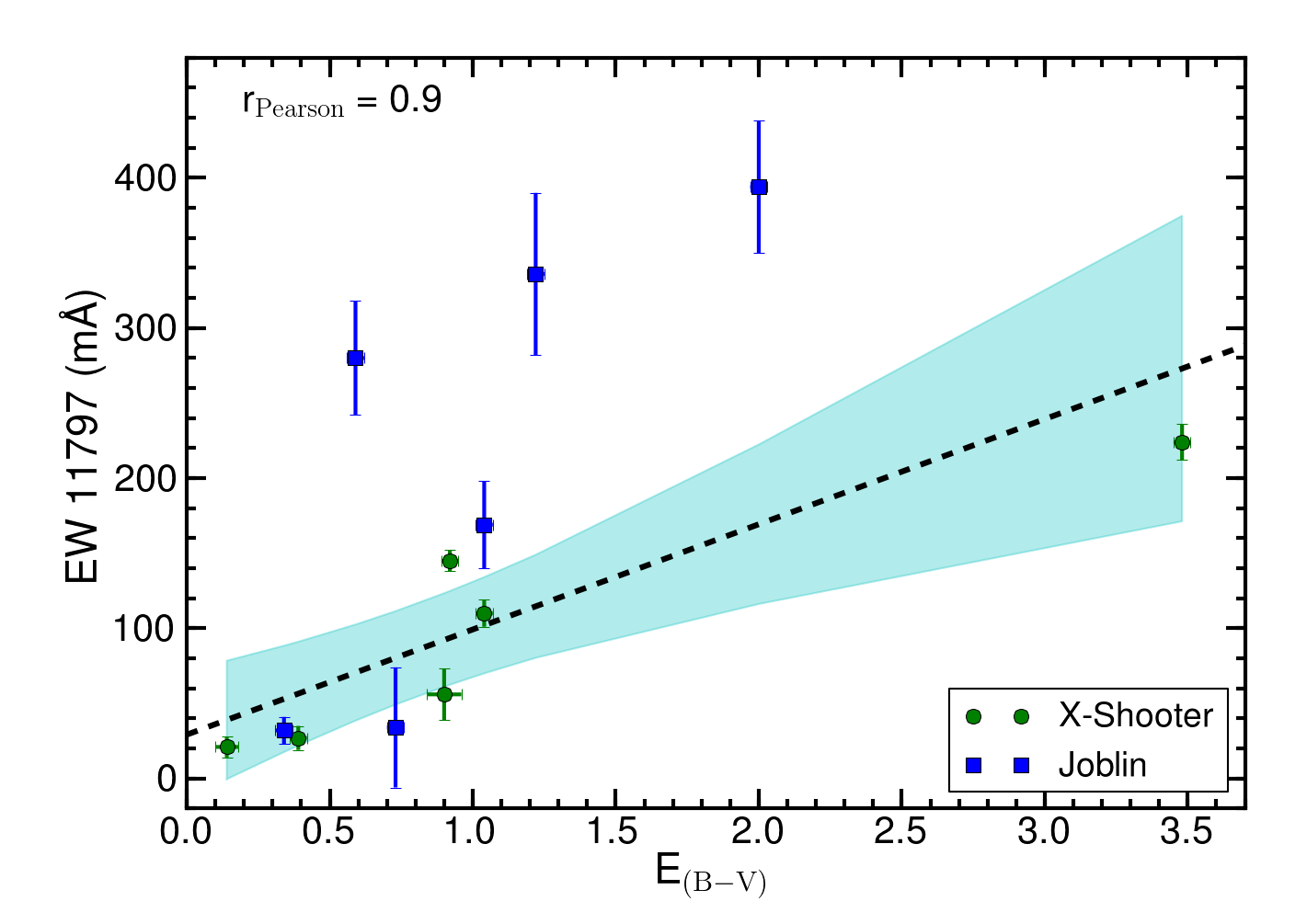}}
\resizebox{0.45\hsize}{!}{\includegraphics{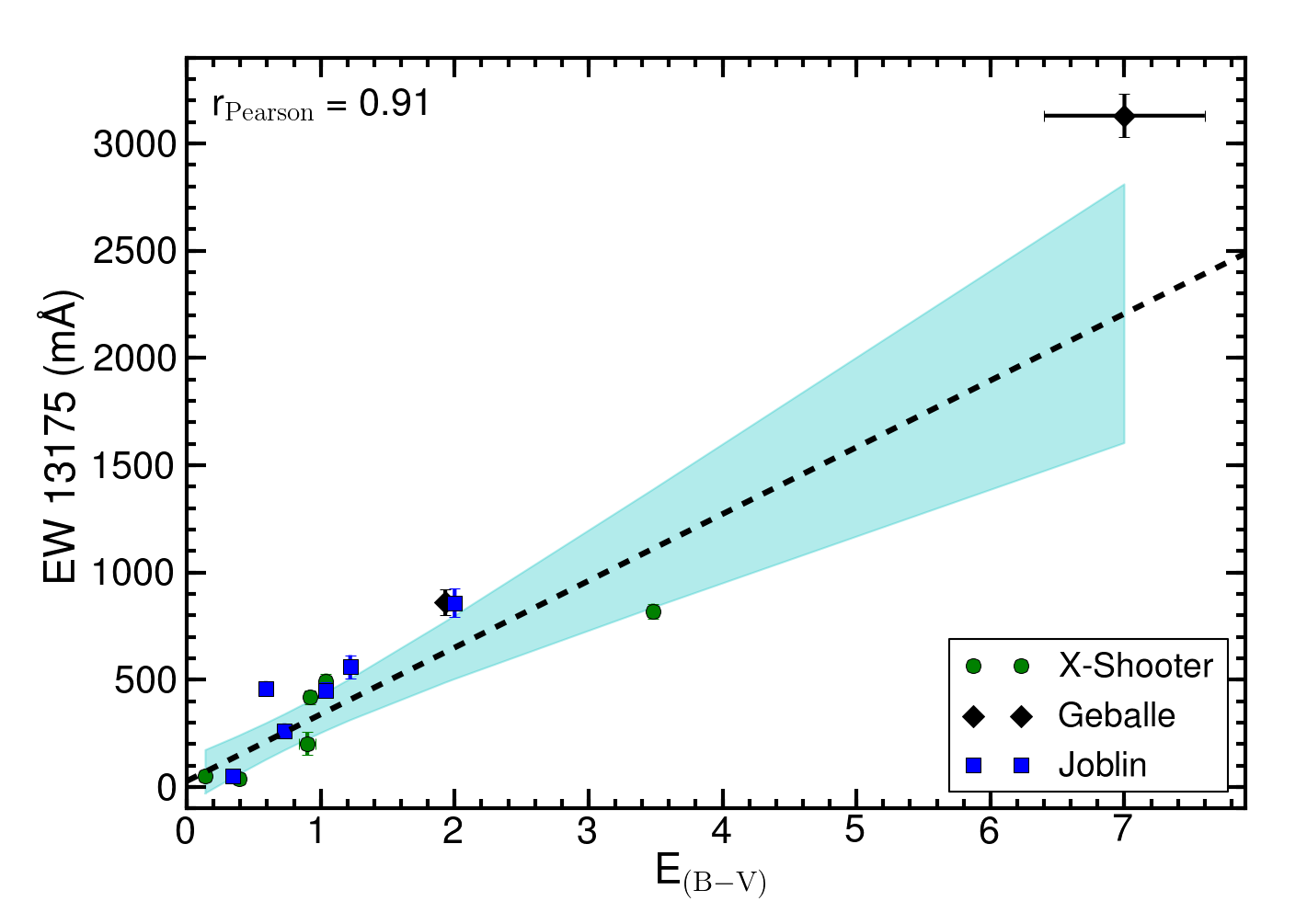}}\\
\resizebox{0.45\hsize}{!}{\includegraphics{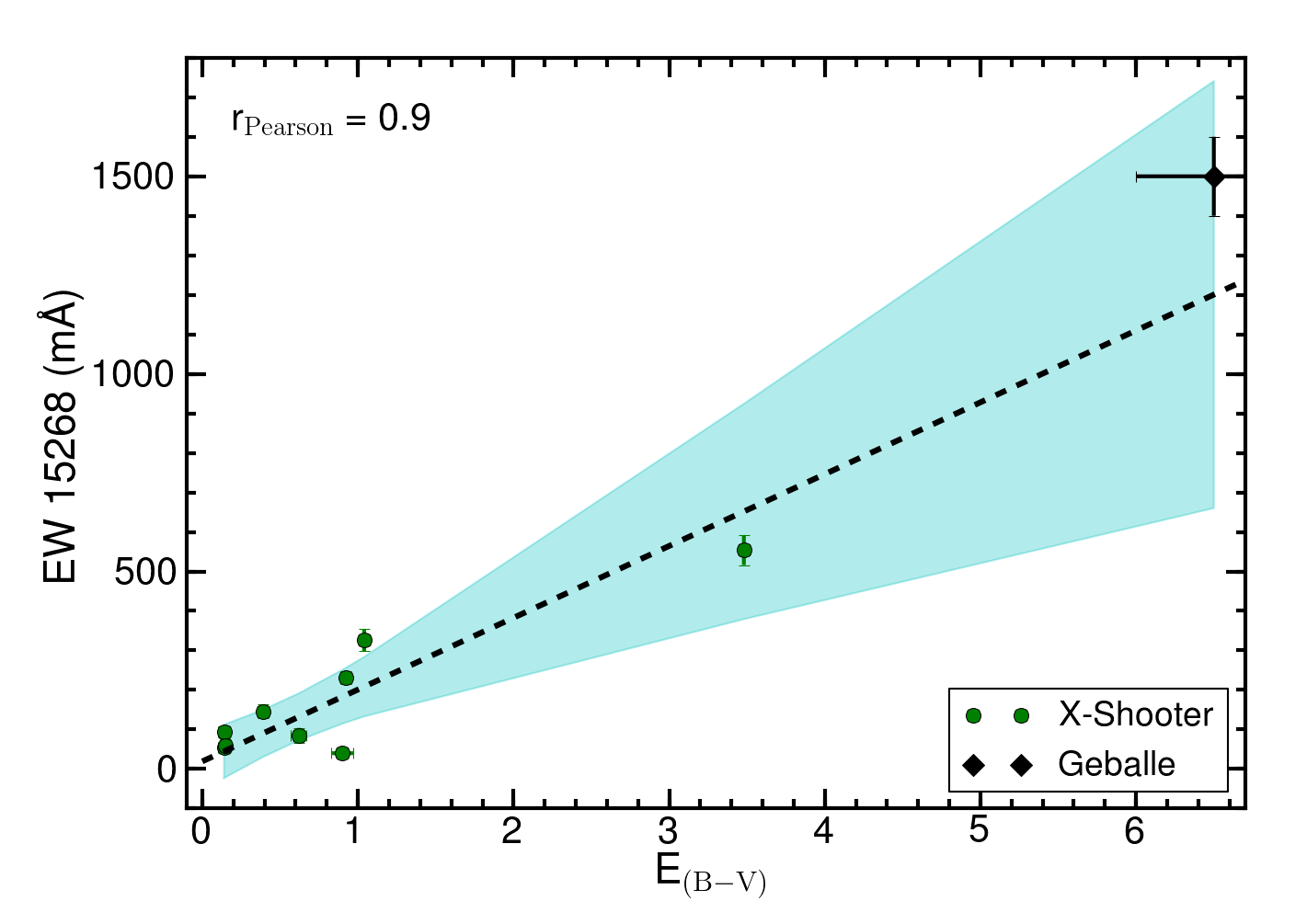}}
\resizebox{0.45\hsize}{!}{\includegraphics{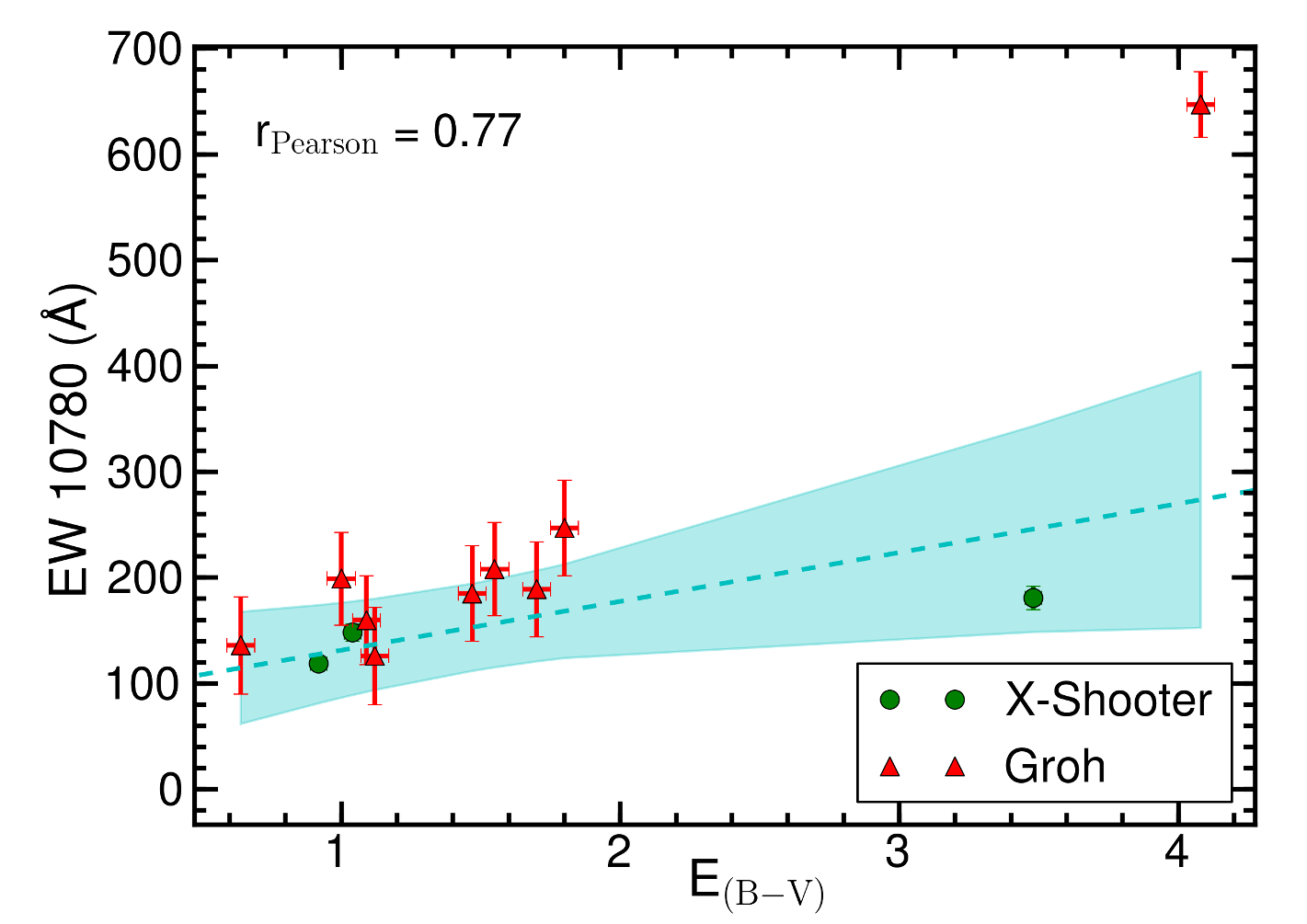}}
 \caption{$\lambda\lambda$11797, 13175, 15268, and 10780 NIR DIB equivalent widths (\AA) as a function of reddening, \Ebv including
   measurements from \citet[][blue squares]{1990Natur.346..729J}, \citet[][see also Table~\ref{tb:nirdib_measurements}; black
   diamonds]{2011Natur.479..200G}, \citet[][red triangles]{2007A&A...465..993G} and this work (green circles). 
   The linear least-square fit is indicated by the black dashed line (parameters are given in Table~\ref{tb:r_pearson})
   and the shaded cyan area indicates the 95\% confidence interval.
 }
 \label{fig:NIRDIBs-Ebv}
\end{figure*}

Fig.~\ref{fig:NIRDIBs-Ebv} shows the measured equivalent widths for the $\lambda\lambda$11797, 13175, and 15268 NIR DIBs as a 
function of reddening, and linear regression fits to the data. For the linear regression the 95\% confidence interval is also 
indicated (shaded area). The corresponding correlation and linear fit coefficients are given in Table~\ref{tb:r_pearson}. 
Notacibly, the strength of the 11797~\AA\ NIR DIB falls well above the linear trend for three lines of sight reported in 
\citet{1990Natur.346..729J}. At this time we cannot confirm if any systematics (such as telluric residuals) affect these 
measurements or if they are truly peculiar, even though for the only sightline in common, HD\,183143, we report a $\sim$35\% 
smaller equivalent width. Although the sample is limited, we find for most NIR DIBs a clear, but imperfect, trend of 
increasing DIB strength with increasing \Ebv. 

\subsubsection{DIB families}

Table~\ref{tb:r_pearson} also lists the correlation coefficients between the three selected NIR DIBs and the optical DIBs 
(for consistency and avoiding systematics these correlation coefficient only take the X-shooter data into account. 

Correlation coefficients are $r>0.8$ between the 15268~\AA\ NIR DIB and the optical DIBs. 
The optical DIBs we consider are also well correlated with each other, $r \geq 0.9$.  
All correlations between the 15268~\AA\ DIB and the optical DIBs are shown in Fig.~\ref{fig:DIBcorrelation-appendix}. 
The 15268~\AA\ NIR DIB is only moderately correlated ($r \sim 0.9$) with both the 11797~\AA\ or 13175~\AA\ NIR DIBs. 
The latter two are strongly correlated with each other, $r = 0.97$.

Since the number of data points is small (between six and nine) and there is significant scatter on the derived linear 
regressions we caution against over-interpretation of the correlation coefficients presented here. 
A larger sample is required for each diffuse band to substantiate or invalidate these preliminary relations, which could be
nevertheless useful in, for example, preparing follow-up observations.
Nonetheless, our measurements reveal that optical and NIR DIBs display a similar behaviour with respect to interstellar 
physical conditions as probed by different lines of sight. Furthermore, the data tentatively suggest that the NIR DIB at 
15268~\AA\ NIR DIB is moderately well correlated with all the optical DIBs. The weakest correlation is found
for the optical DIBs at 5797 and 5849~\AA. 
The relation between NIR DIBs and the 5780~\AA\ DIB (which is considered as a good tracer of neutral hydrogen; 
\citealt{1993ApJ...407..142H}) is discussed in Sect.~\ref{sec:hd147889}. Considering the small sample size we do
not consider further subdivision of the sample according to, for example, $\sigma$/$\zeta$-type, etcetera.

\begin{figure}[tp!]
\centering
\resizebox{\hsize}{!}{\includegraphics{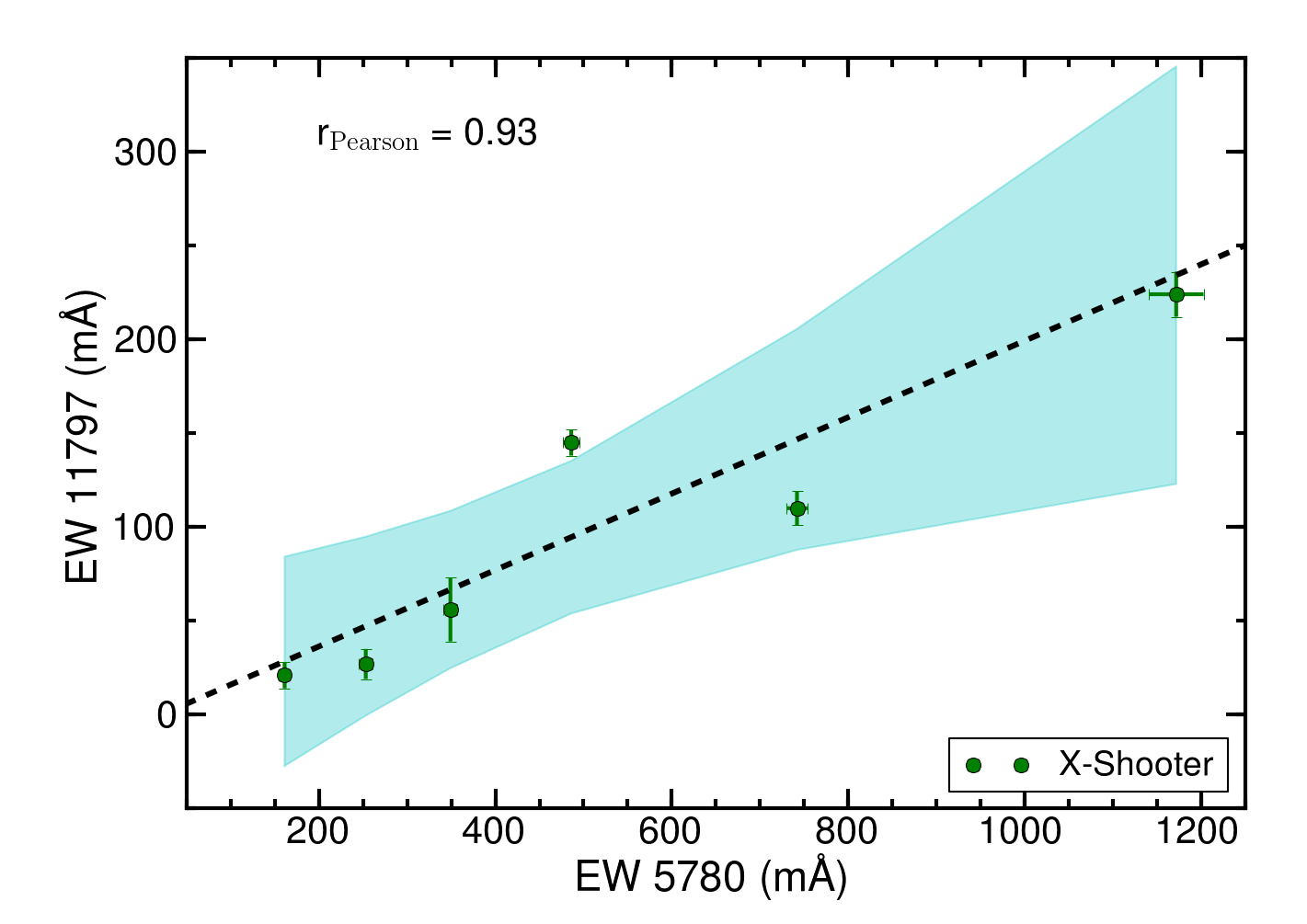}}
\resizebox{\hsize}{!}{\includegraphics{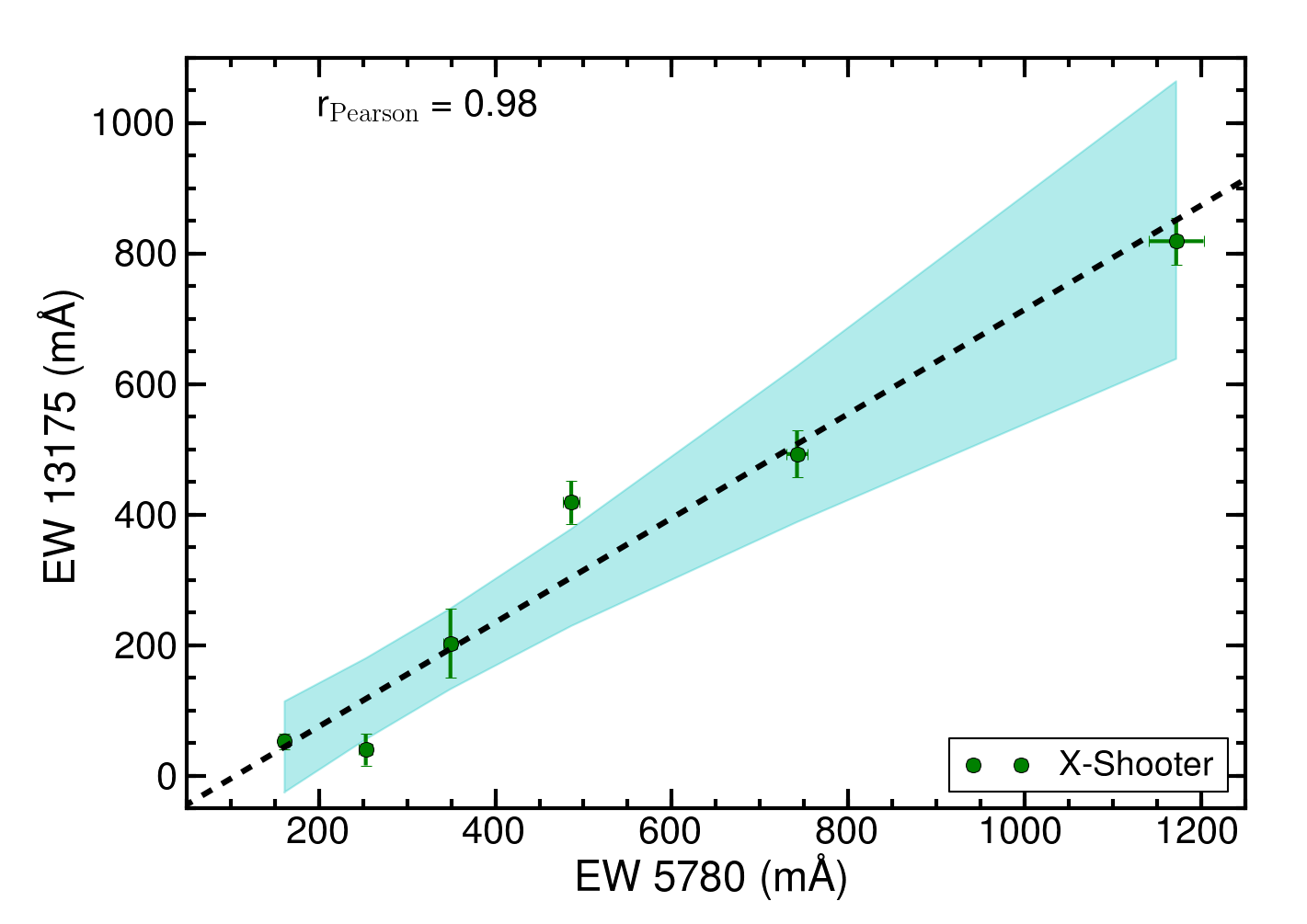}}
\resizebox{\hsize}{!}{\includegraphics{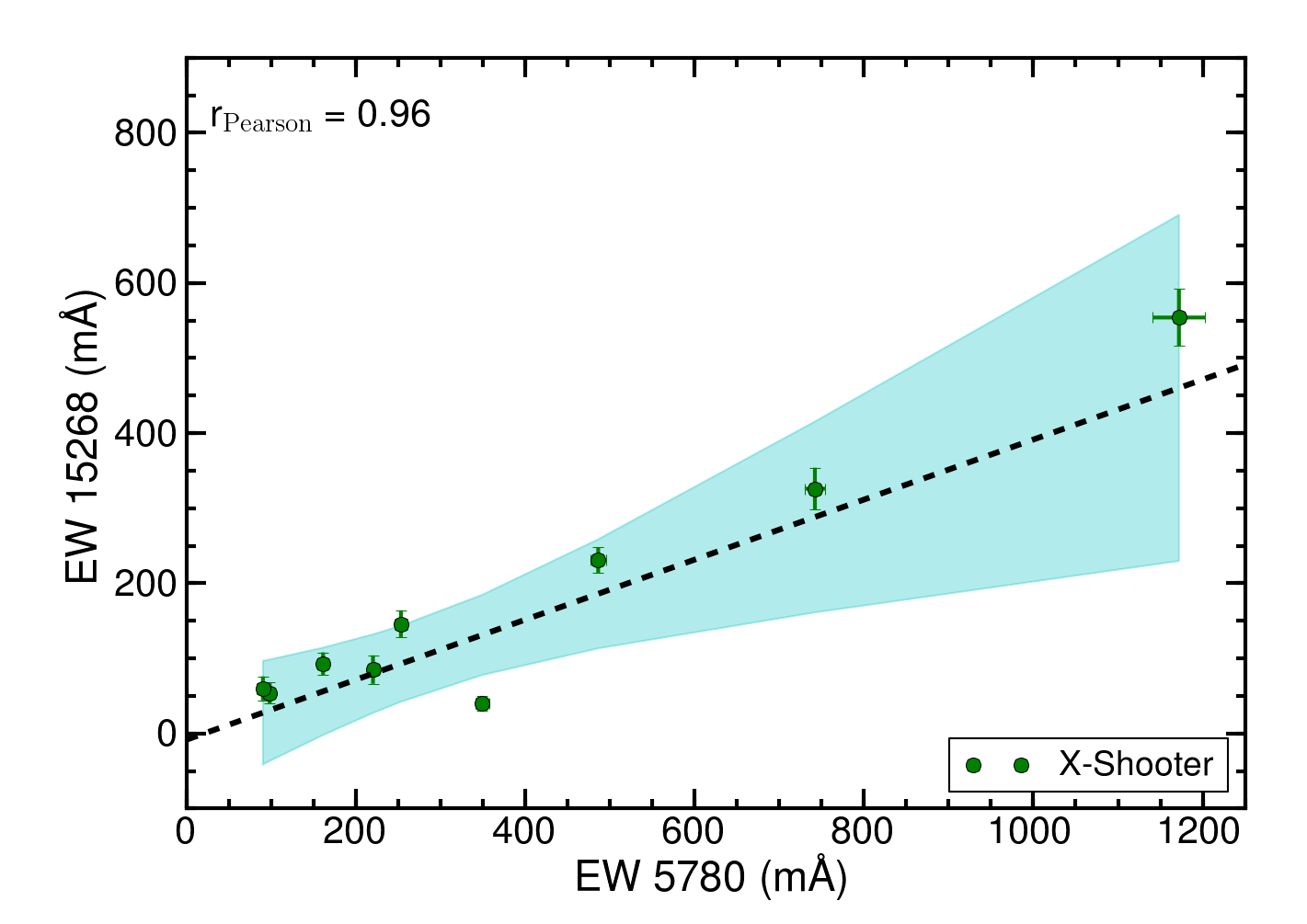}}
\caption{Mutual correlation between the 5780~\AA\ DIB and the $\lambda\lambda$ 11797, 13175, and 15268~NIR DIBs.
 The correlation coefficients are given in each panel.
 The linear least-square fit is indicated by the black dashed line (parameters are given in Table~\ref{tb:r_pearson})
 and the shaded cyan area indicates the 95\% confidence interval.} 
 \label{fig:5780DIB-NIRDIB}
\end{figure}

\begin{table}[ht!]
\centering
\caption{Correlation coefficients $r$ and best-fit parameters of the linear regression $y = a*x + b$ for the listed parameter 
pairs for the X-shooter data. Equivalent widths, W, given in m\AA.}
\label{tb:r_pearson}
\begin{tabular}{r@{-}llll}\hline\hline
\multicolumn{2}{c}{parameter pair $y-x$\tablefootmark{a}} & $r$   & $a$\tablefootmark{b} & $b$ (m\AA) \tablefootmark{b} \\ \hline
W(11797)  & \Ebv      & 0.90	  & 70 (17)     &  29 (19) \\
W(13175)  & \Ebv      & 0.91	  & 311 (44)    &  28 (50) \\
W(10780)  & \Ebv      & 0.77	  & 46 (19)     &  85 (33) \\
W(15268)  & \Ebv      & 0.90	  & 182 (40)    &  19 (33) \\ \vspace{3mm}
W(13175)  & W(11797)  & 0.97	  & 3.76 (0.42)	& 37 (43) \\
W(11797)  & W(5780)   & 0.92	  & 0.20 (0.55)	& -4 (29) \\
W(13175)  & W(5780)   & 0.98	  & 0.80 (0.91)	& -84 (37)\\ \vspace{3mm}
W(15268)  & W(5780)   & 0.96	  & 0.40 (0.11)	& -8 (37) \\
W(15268)  & W(5797)   & 0.91	  & 0.87 (0.35)	& 14 (44) \\
W(15268)  & W(5849)   & 0.83	  & 1.40 (1.02)	& 43 (56) \\
W(15268)  & W(6196)   & 0.95	  & 3.20 (0.96)	& -4 (36) \\
W(15268)  & W(6203)   & 0.97	  & 1.32 (0.24)	& 4 (24)  \\
W(15268)  & W(6270)   & 0.97	  & 1.53 (0.26)	& 24 (21) \\
W(15268)  & W(6613)   & 0.95	  & 0.74 (0.22)	& -2 (39) \\
W(15268)  & W(11797)  & 0.90	  & 1.98 (0.56)	& 31 (54) \\
W(15268)  & W(13175)  & 0.93	  & 0.57 (0.13)	& 48 (39) \\ \hline
\end{tabular}
\tablefoot{
\tablefoottext{a}{Equivalent width of the NIR DIB at the given wavelength or interstellar reddening.}
\tablefoottext{b}{The linear regression coefficients including uncertainties as shown in 
Figs.~\ref{fig:NIRDIBs-Ebv}, \ref{fig:5780DIB-NIRDIB}, and~\ref{fig:DIBcorrelation-appendix}.}
}
\end{table}

\begin{figure*}[bh!]
\centering
\resizebox{.745\hsize}{!}{
 \includegraphics{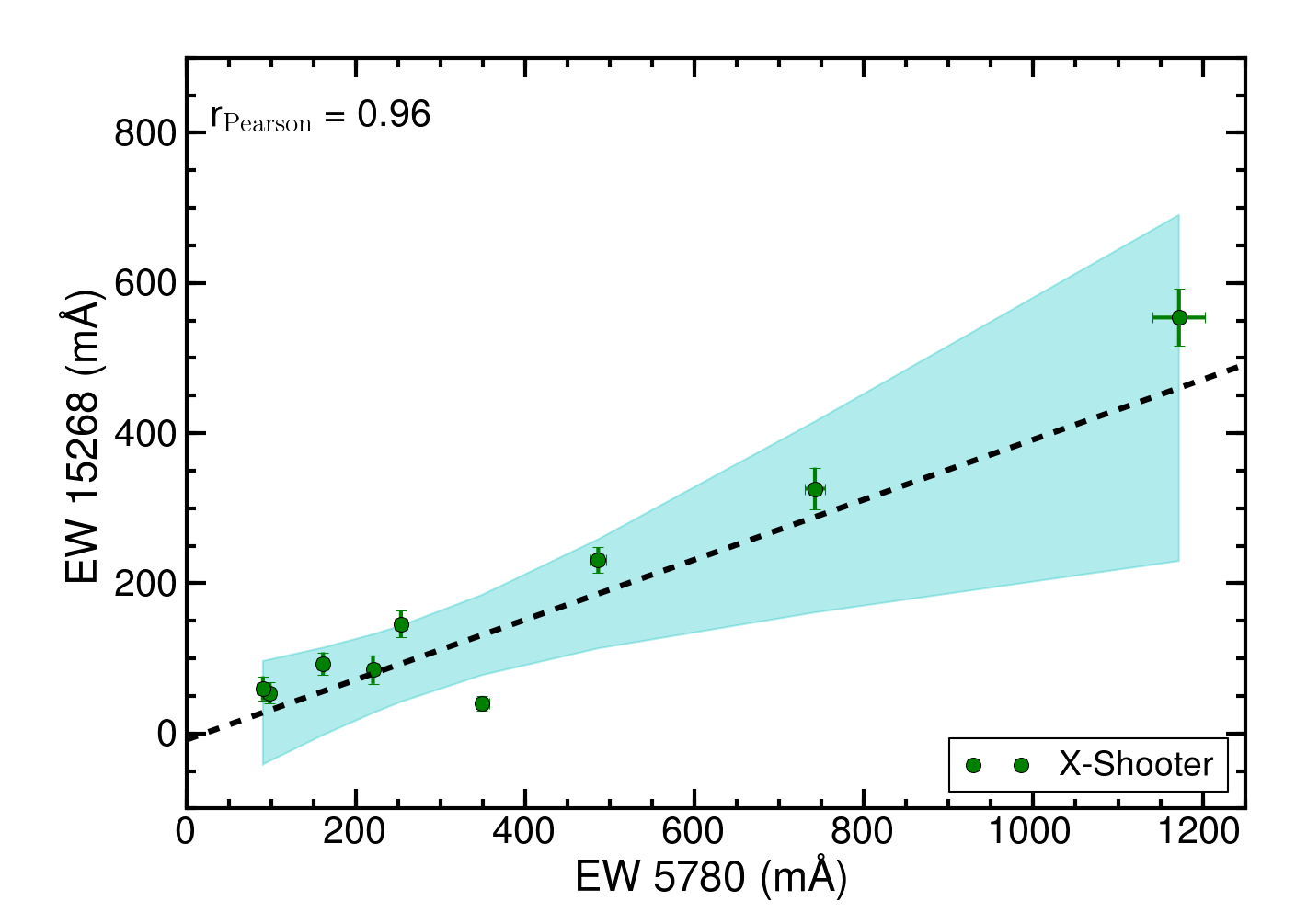} \hspace{20mm} \includegraphics{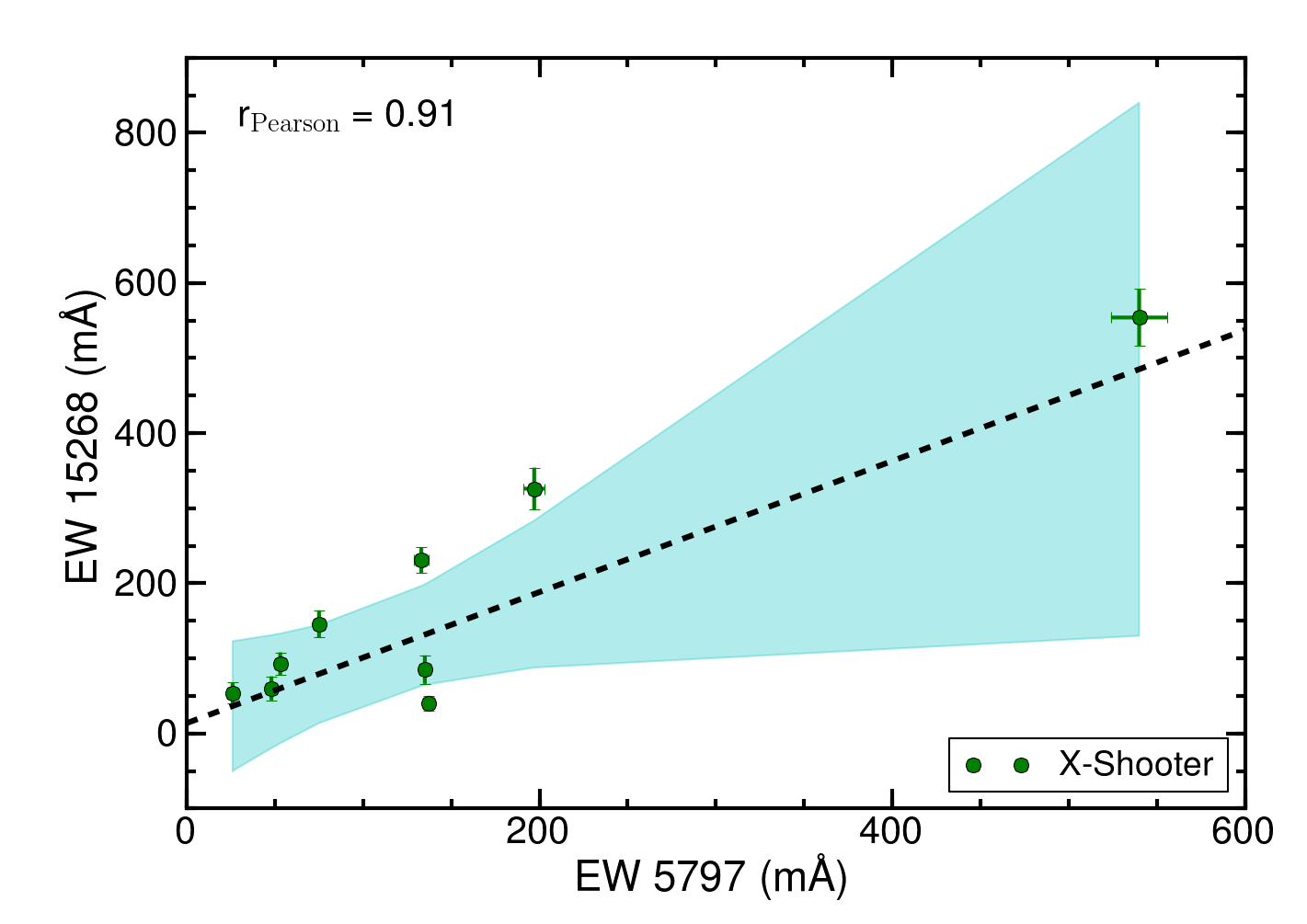}}

\resizebox{.745\hsize}{!}{ 
 \includegraphics{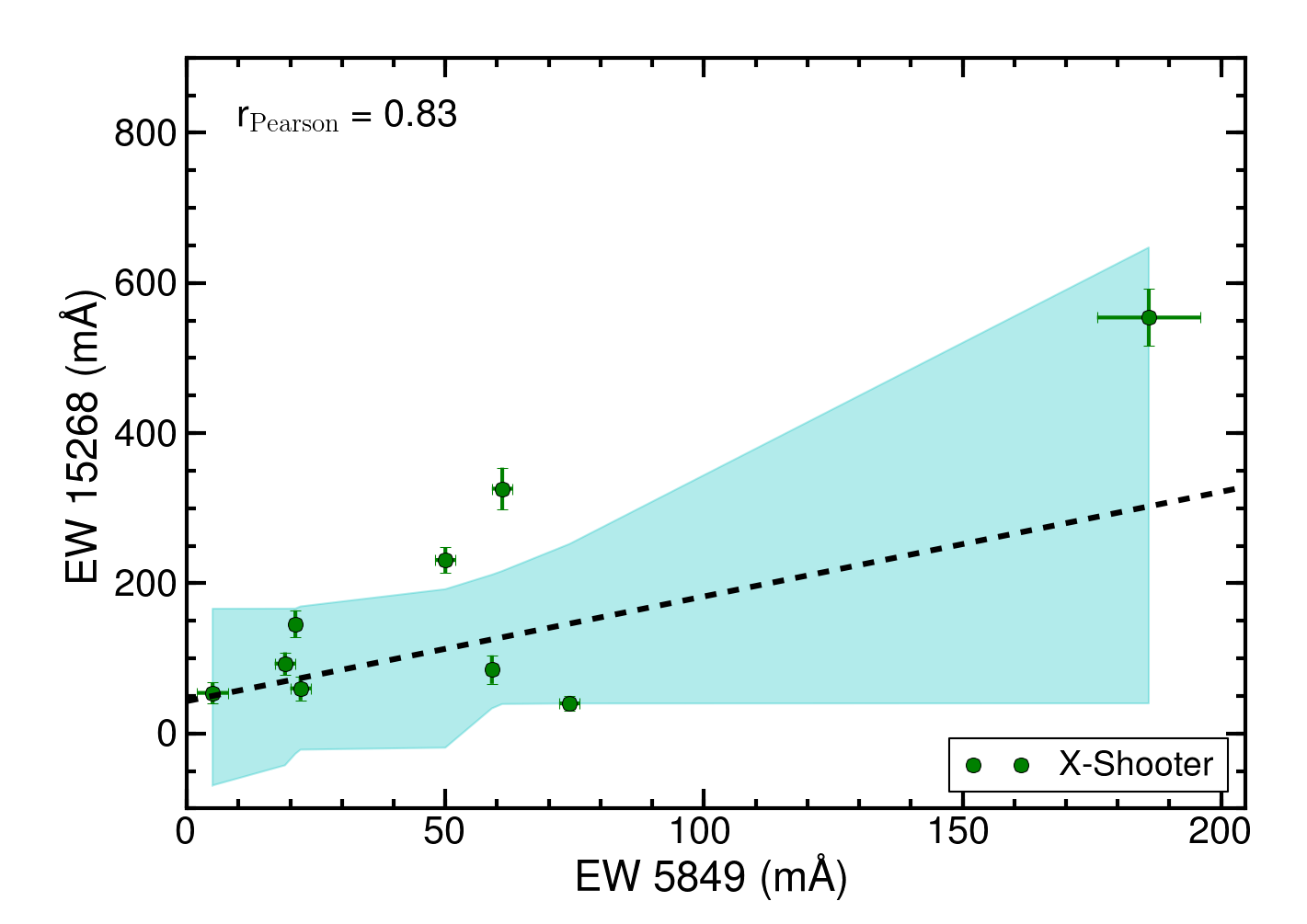} \hspace{20mm} \includegraphics{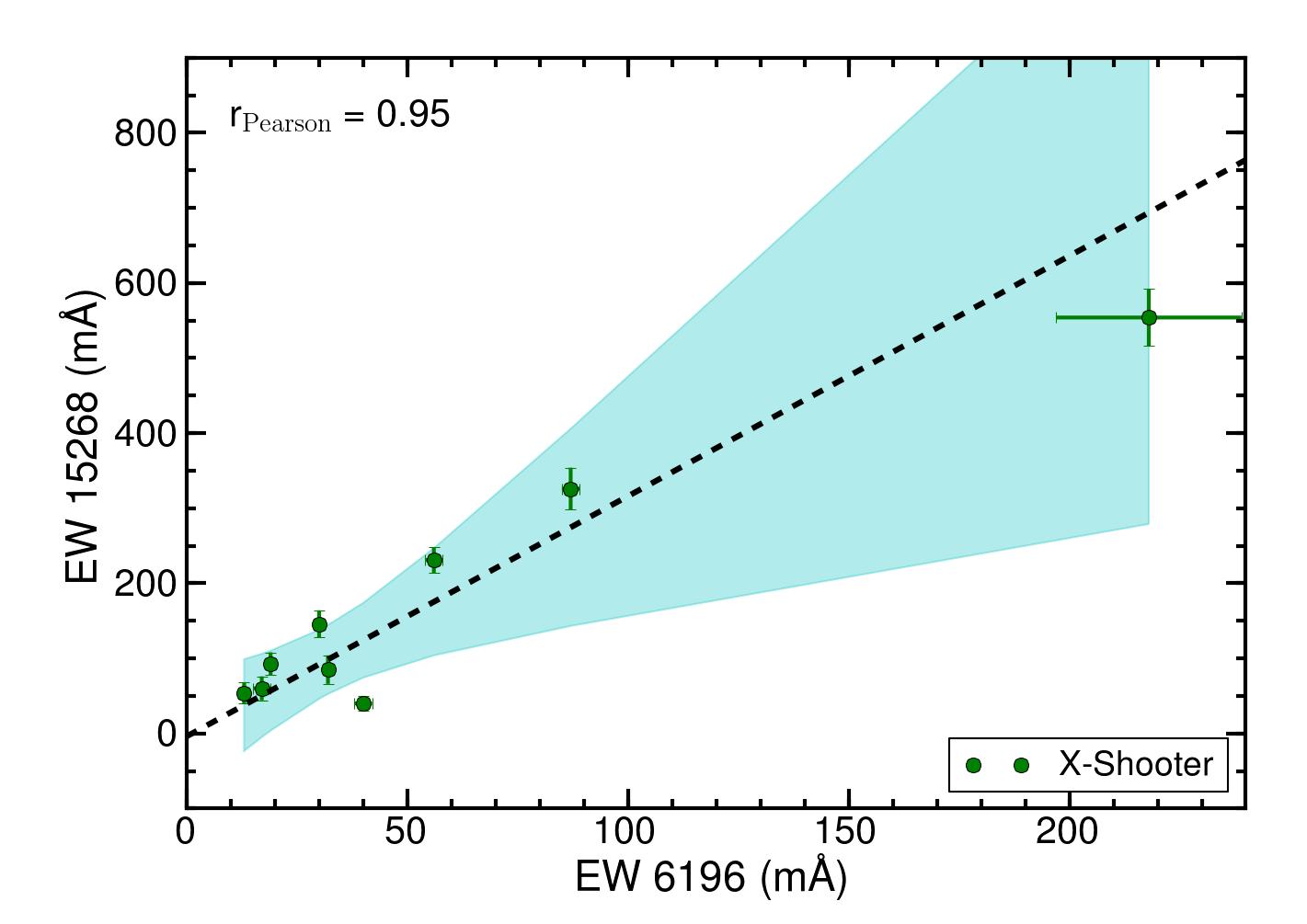}}

\resizebox{.745\hsize}{!}{
 \includegraphics{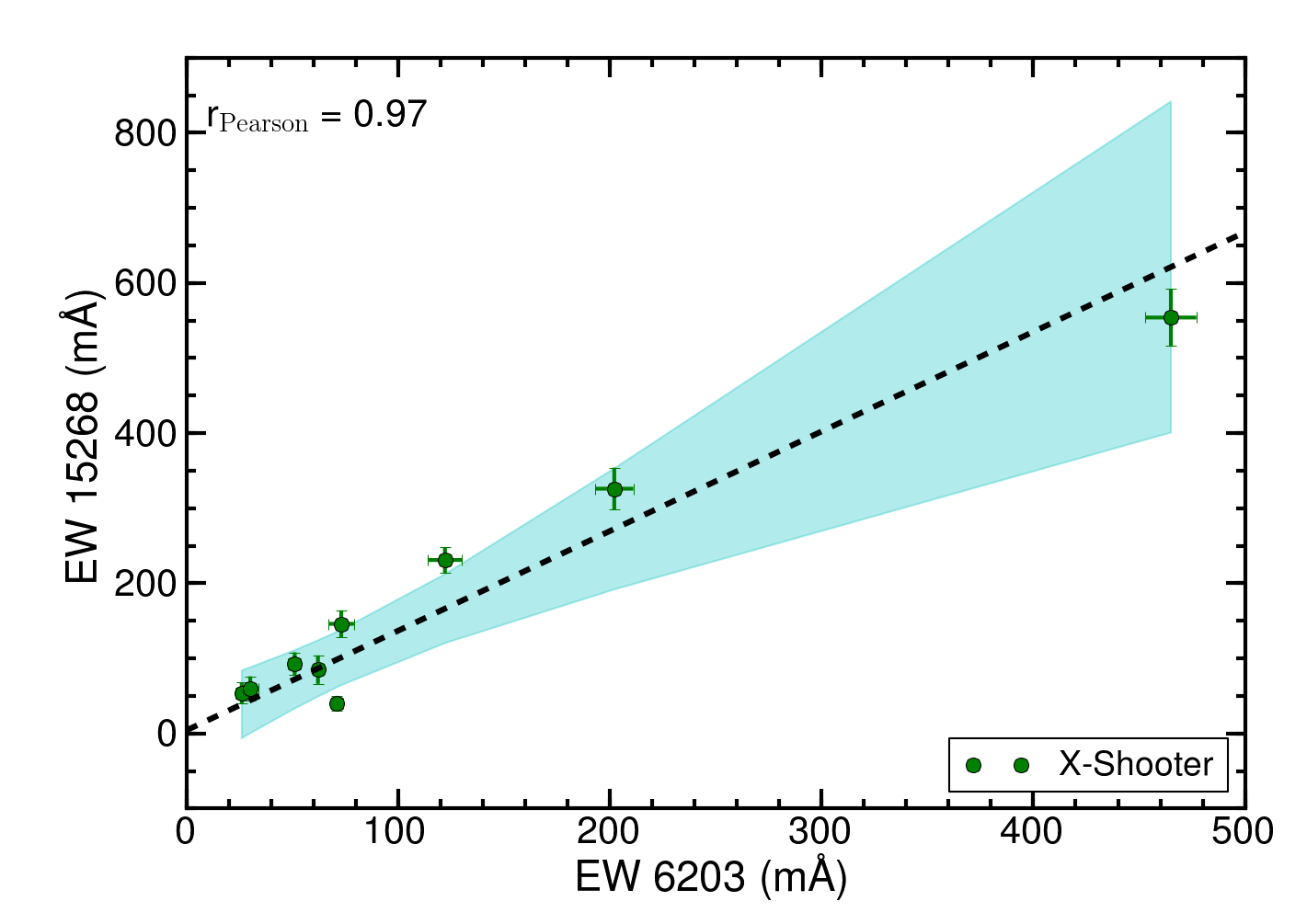}  \hspace{20mm} \includegraphics{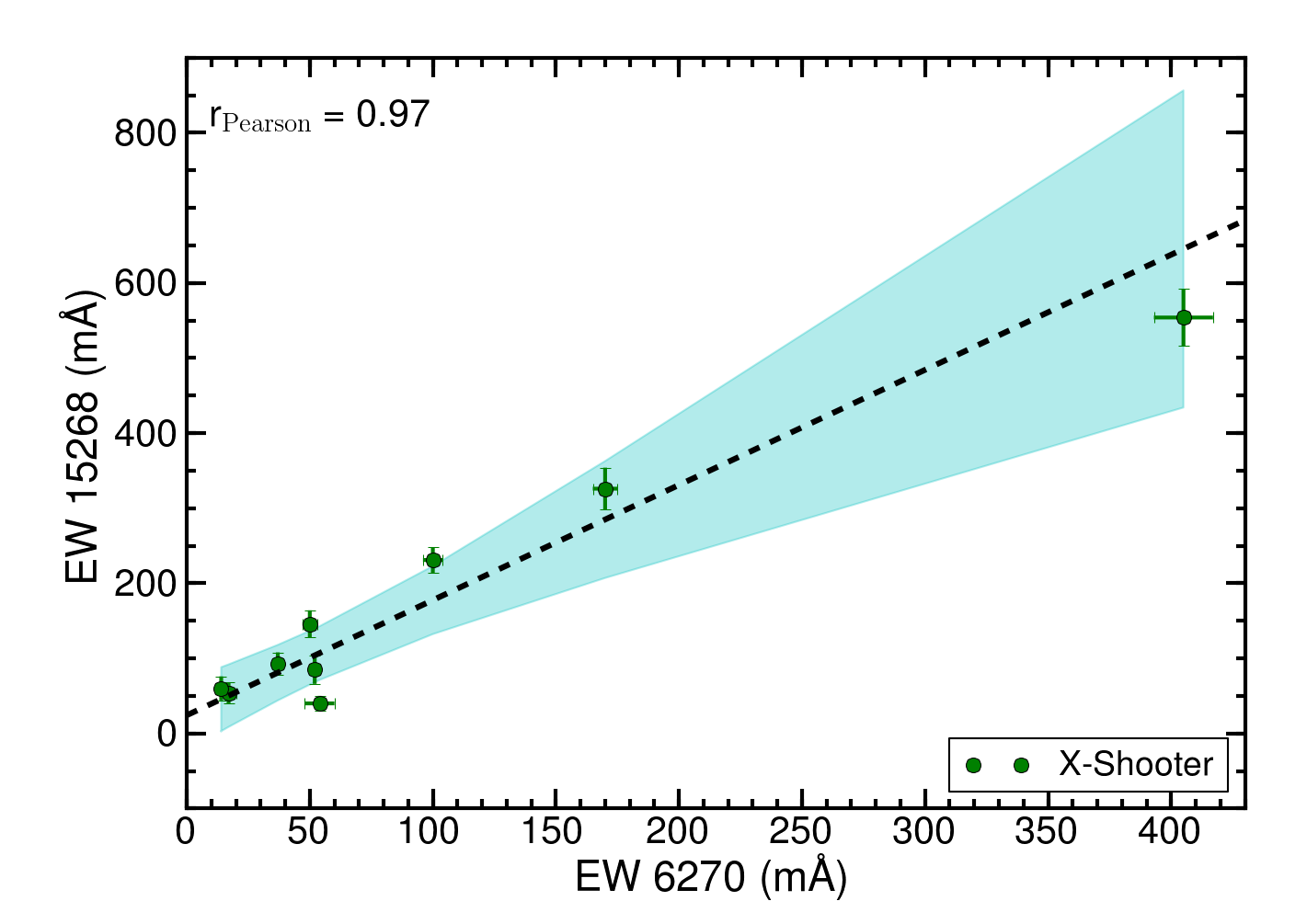}}

\resizebox{.745\hsize}{!}{
 \includegraphics{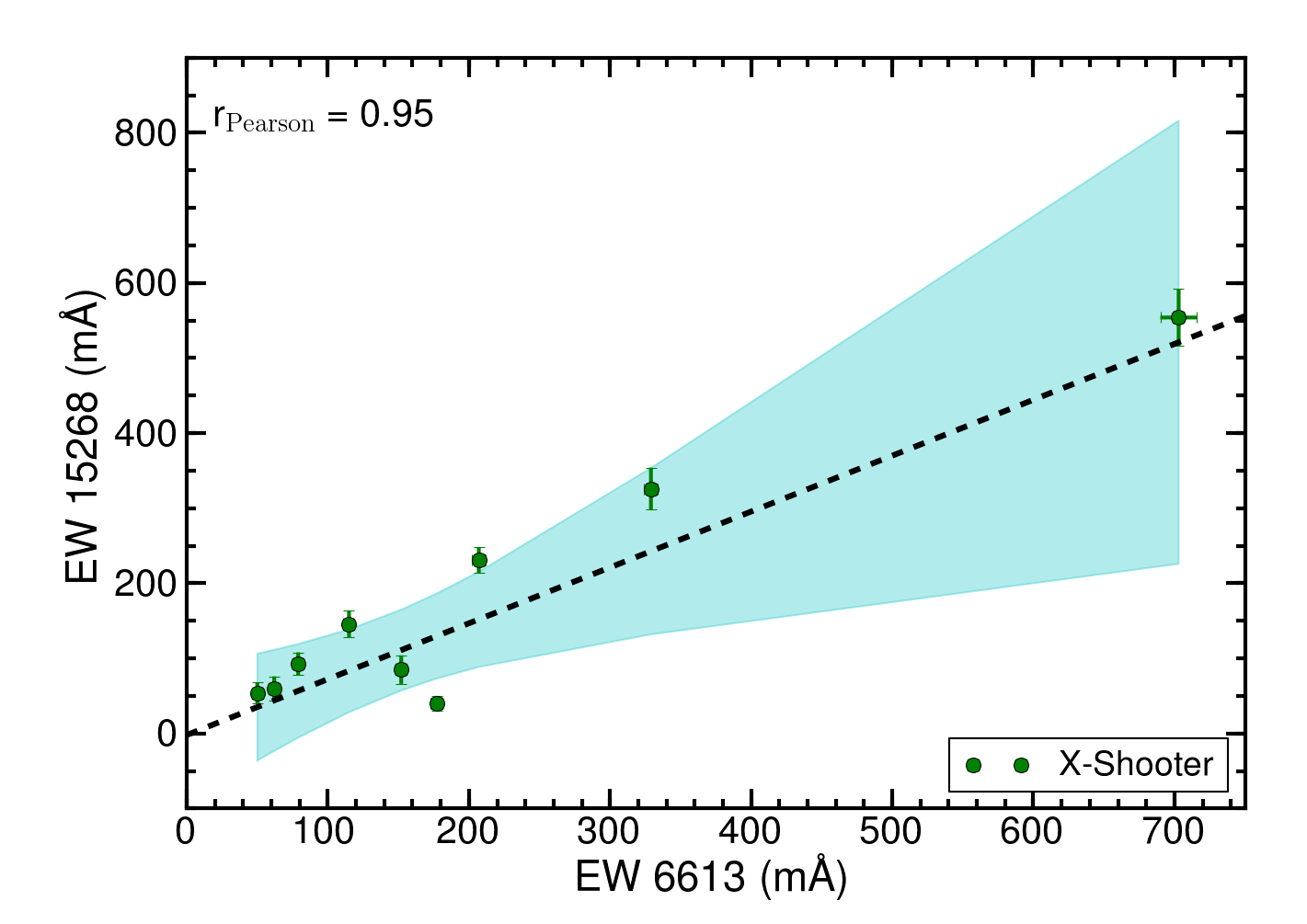} \hspace{20mm} \includegraphics{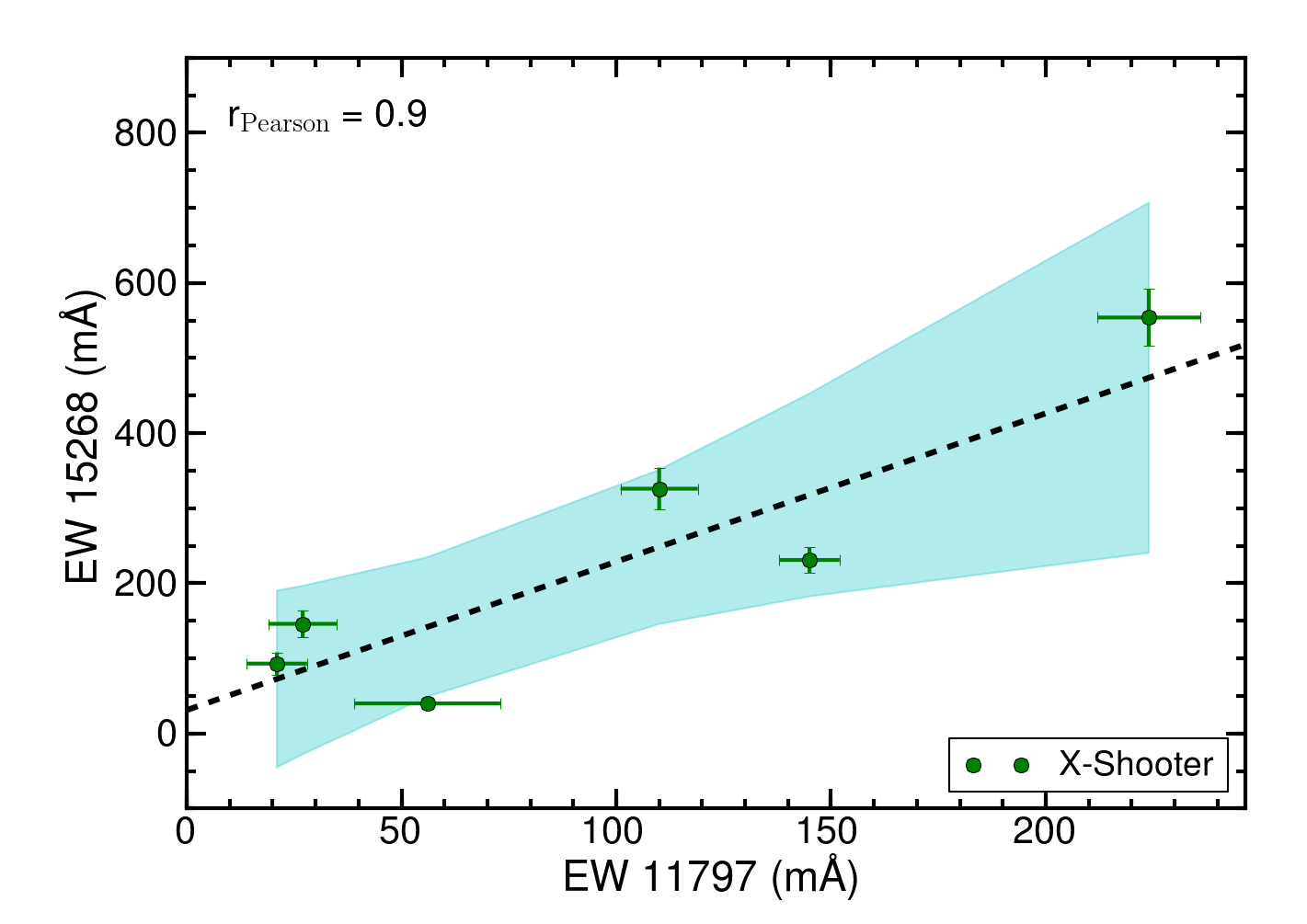}}
 
\resizebox{.745\hsize}{!}{
 \includegraphics{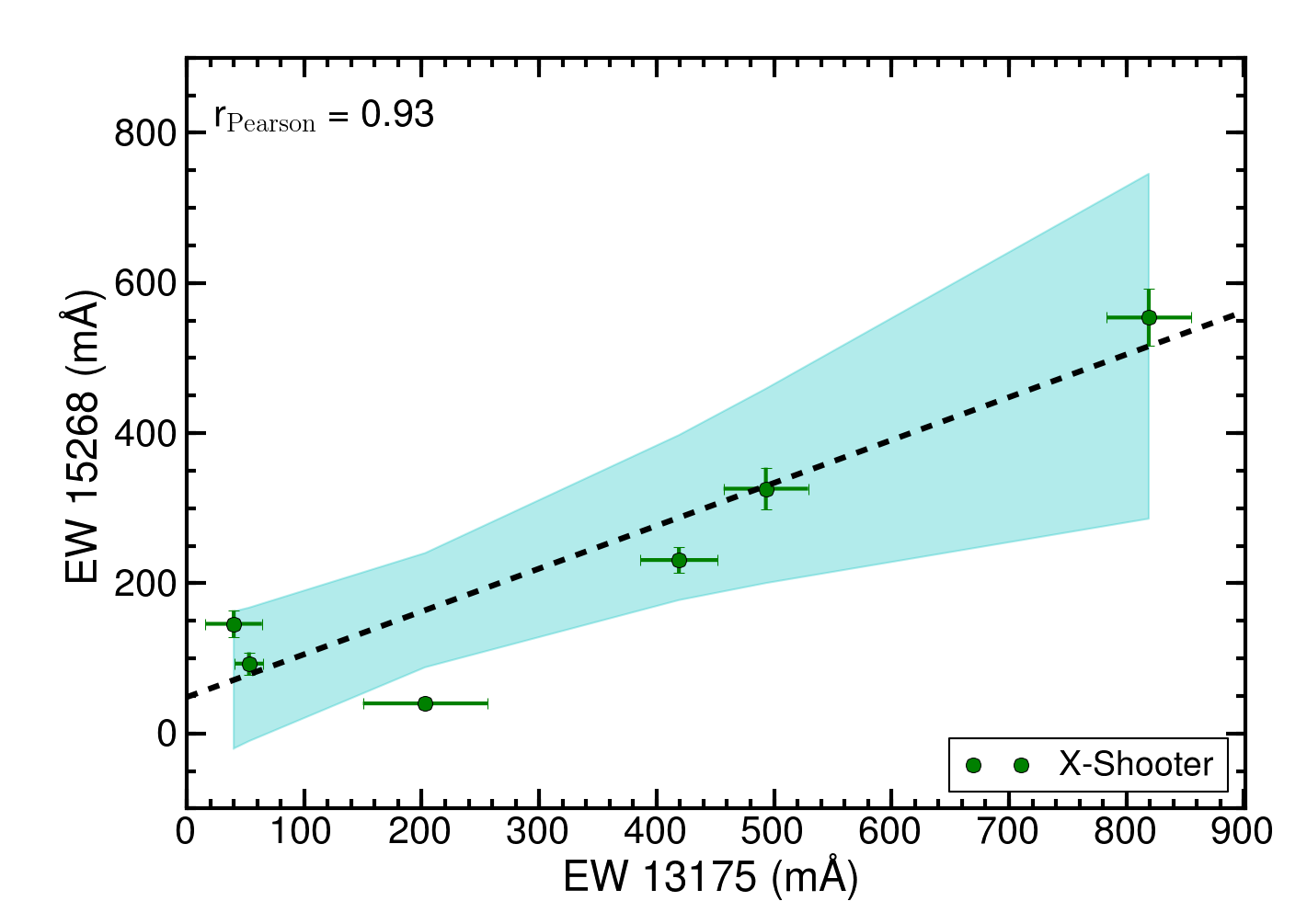} \hspace{20mm} \includegraphics{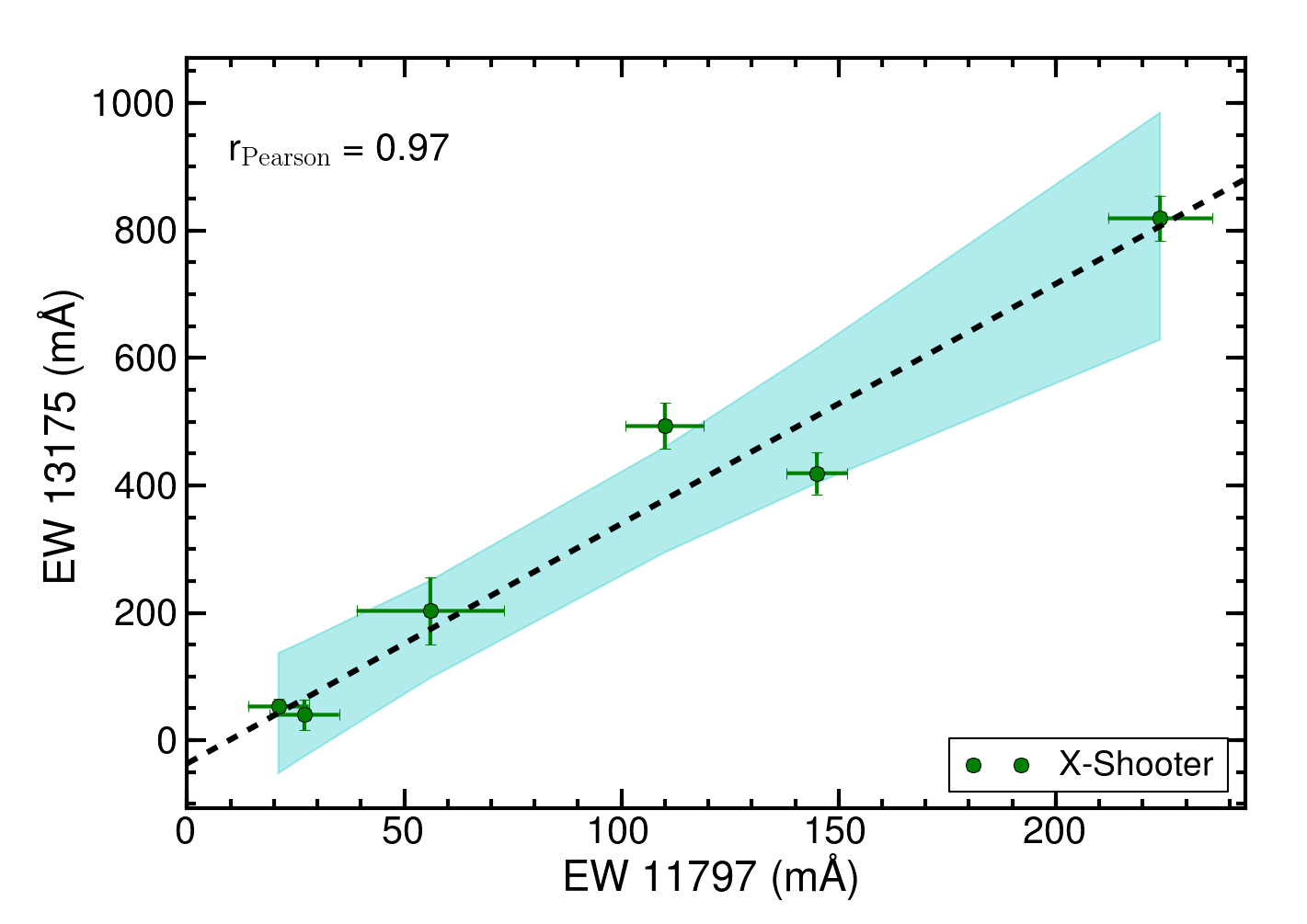}}

 \caption{Relation between the 15268~\AA\ NIR DIB and the strong optical DIBs, and the 11797 and 13175~\AA\ NIR DIBs. 
 The correlation coefficients are given in each panel. 
 In addition the 11797-13175 NIR DIBs are compared in the bottom right panel. 
 The linear least-square fit is indicated by the black dashed line (parameters are given in Table~\ref{tb:r_pearson})
 and the shaded cyan area indicate the 95\% confidence interval.
 }
 \label{fig:DIBcorrelation-appendix}
\end{figure*}

\section{NIR DIBs and large PAH cation bands}\label{sec:laboratory}

In the context of the DIB-PAH proposal, the NIR range could be particularly interesting since many PAH ions have strong electronic
transitions in this wavelength range \citep{2005ApJ...629.1188M}. Specifically, these authors measured the spectra of 27 PAH cations 
and anions, ranging in size from C$_{14}$H$_{10}$ to C$_{50}$H$_{22}$, and found that they have strong and broad absorption bands 
between 0.7 and 2.5~$\mu$m. Smaller PAHs on the other hand have their strongest transitions in the optical or UV. 
Since it is thought that the population of interstellar PAHs is dominated by species in the $N_C=50$ range (see
e.g. \citealt{2008ARA&A..46..289T}), the NIR is ideally suited to study and possibly identify these species in space. 

The PAH ion absorption spectra presented by \citet{2005ApJ...629.1188M} are typically characterised by one or two
strong transitions accompanied by several weaker transitions. Ten of the PAH cations presented have bands with 
oscillator strengths $f \leq 0.02$ longwards of 1.0~$\mu$m. Here, we make an initial comparison between these laboratory 
measurements and the NIR DIBs. Note however that the absorption band wavelengths reported by \citet{2005ApJ...629.1188M} 
result from matrix-isolation experiments and hence suffer from an unknown shift in position (by as much as tens
of Angstrom) as well as from band broadening due to interaction with the solid lattice. 
Thus, while a direct identification of PAH cations is not possible with this data set, a comparison between the
laboratory data and the NIR DIBs can offer some insights into this population of carbonaceous species. 
Here, we present some of the more interesting cases; the numbering refers to that used by \citet{2005ApJ...629.1188M}: 

\begin{enumerate}

\item Molecule no. 10 (C$_{20}$H$_{12}^+$) has two strong bands at 10510~\AA\ ($f = 0.077$) and 9277~\AA\ ($f = 0.008$).  
We identified two new candidate DIBs at 10504~\AA\ and 10506~\AA, but could not verify the presence of the second, 
weaker band near 9277~\AA\ due to a strong telluric line forest in this range.

\item Molecule no. 16 (C$_{36}$H$_{16}^+$) shows a strong laboratory band at 10520~\AA~($f = 0.031$), with 
two weaker bands (by one order of magnitude) at 10220~\AA\ and 9196~\AA. Again, two tentative NIR DIBs are identified at
10504~\AA\ and 10506~\AA. No feature is revealed at 10220~\AA, but a shallow broad feature is suspected at 9212~\AA.

\item Molecule no. 18 (C$_{40}$H$_{18}^+$) has a strong band at 10410~\AA~($f = 0.024$), near the new candidate at
10438~\AA. The weaker bands at 9789~\AA\ and 8918~\AA~(laboratory) could not be discerned in the X-shooter spectra.

\item Molecule no. 19 (C$_{40}$H$_{18}^+$) has a strong band at 13010~\AA~($f = 0.026$), near the new candidate at
13026~\AA. Secondary laboratory bands with $f \sim 0.002$ are present at 9233, 7854, and 7693~\AA. 
As noted above, a near infrared band may be present at 9212~\AA. Furthermore, strong DIBs have been
identified at 7686.5, 7705.9, and 7710~\AA~\citep{1994A&AS..106...39J}. 

\item Molecule no. 22 (C$_{42}$H$_{22}^+$) has a strong band at 10790~\AA~($f = 0.042$), near the new candidate at
10780~\AA. The secondary bands at 9386 and 7209~\AA~($f \sim 0.004$) are not found.

\item Molecule no. 25 (C$_{48}$H$_{20}^+$) has a strong band at 10410~\AA~($f = 0.089$), near the new candidates 
at 10392~\AA\ and 10438~\AA. Its strongest feature is at 6941~\AA~($f = 0.160$) which is near a strong broad DIB 
at 6939~\AA\ with FWHM of 21.3~\AA.

\end{enumerate}

We do not find any features in the vicinity of the main transitions reported for molecules  no. 23
(C$_{44}$H$_{20}^+$; $f = 0.025$), no. 24 (C$_{48}$H$_{20}^+$; $f = 0.061$), no. 26 (C$_{48}$H$_{22}^+$; $f = 0.031$).

The above comparison shows several coincidences between the newly detected NIR DIBs and the laboratory spectra of PAH cations, 
thus illustrating the potential of this low-congested range of the DIB spectrum. 
However, the measured PAH cation bands are much broader (at least tens of \AA) than the rather narrow NIR DIBs. 
While part of the measured bandwidth for the PAHs certainly results from matrix broadening effects, 
it is clear that PAH cation bands are intrinsically broad as well, and thus possibly too broad to match the NIR DIBs. 
A conclusive answer can therefore only come from precise transition wavelengths and information on the intrinsic band profiles 
and widths, obtained from gas-phase spectra of possible carrier molecules as has been illustrated for DIBs in the optical
range \citep[see e.g.][]{2011A&A...530A..26G,2011ApJ...728..154S}.

\section{The translucent cloud toward HD\,147889}\label{sec:hd147889}

In Sect.~\ref{sec:NIRDIBs} we found that many of the NIR DIBs towards HD\,147889 are weak compared to sightlines with similar reddening.
On the other hand, strengths of the optical DIBs are consistent with those found in similar lines of sight. 
For instance the optical DIBs toward HD\,161061, HD\,147889 and HD\,183143 have similar strengths per unit reddening, 
while the NIR DIBs at 9577, 9632, and 15628~\AA\ are weaker. The strengths of the 11797 and 13175~\AA\ NIR DIBs 
toward HD\,147889 are in line with the good linear relation with the 5780~\AA\ DIB strength, while the 15268~\AA\
NIR DIB is unusally weak compared to the general trend (Fig.~\ref{fig:5780DIB-NIRDIB}).

The line of sight toward HD\,147889 sightline passes through a well known and extensively studied translucent cloud with a 
visual extinction of $\sim$4~mag \citep{2005A&A...432..515R}. Its mean hydrogen density and molecular hydrogen 
fraction are high, $n_H = 1200~\pm~500$~cm$^{-3}$ and $f_{H_2} = 0.4$.
The effective local interstellar radiation field strength for this cloud is 11 times as strong as the Galactic average. 
\citet{2005A&A...432..515R} computed the ionisation balance of a set of small to intermediate size PAHs to construct a PAH 
absorption spectrum ranging from the UV to the near-IR. These models showed a non-negligible fraction of anions and a significant 
fraction of cations. The prediction for this line-of-sight was that if DIBs are due to compact and non-compact PAHs homologue series 
(with 10 $<$ C-atoms $<$ 50) then more DIB features should be present longwards of 10\,000~\AA. 
Thus, the absence of NIR DIBs could indicate that for HD\,147889 large non-compact PAHs are absent or altered, 
for example due to substantial hydrogenation, resulting in weak bands. In addition, the ionisation fraction in this translucent 
cloud is less than 30\% for small PAHs with less than 30 carbon atoms, while the anion fraction is over 20\%. 
Thus the absence of the NIR DIBs (and 5849~\AA\ DIB) could be related to the low ionisation fraction of \emph{smaller} PAHs in 
this line-of-sight as compared to that in diffuse clouds. A lower ionisation fraction - basically due to self-shielding - is expected 
for translucent clouds with a high molecular abundance (\citealt{2006ARA&A..44..367S}). 
Hence, also the reduced strength of the 15268~\AA\ NIR DIB with respect to the 5780~\AA\ DIB which is known to trace particularly well
the amount of neutral hydrogen (\citealt{1995ARA&A..33...19H}, \citealt{2011ApJ...727...33F}). Due to the low ionisation potential 
of large ($>$100 C-atoms) PAHs, these would have a large ionisation fraction ($\geq$ 50\%) even in this translucent cloud.

\section{Summary \& Conclusions}\label{sec:conclusion}\label{sec:summary}

This paper presents moderate-resolution optical and NIR spectra, using X-shooter at the VLT, for eight Galactic early-type stars 
with interstellar reddening values (E$_{B-V}$) ranging from 0 to 3.5~magnitudes (or 10~mag visual extinction correspondingly). 
We present and discuss the presence and properties of NIR DIBs in these sightlines. Specifically, we have obtained the following 
results for the presence of diffuse interstellar bands:

\paragraph{Detection of NIR DIBs}

\begin{itemize}

\item We detect nine of the thirteen new NIR DIBs that were reported by \citet{2011Natur.479..200G}, and can confirm their 
      interstellar nature from their general correlation with reddening and from the stationary nature of the rest 
      wavelengths when corrected for the radial velocity of the interstellar cloud in the line-of-sight.

\item We propose seven new NIR DIB candidates based on their detection in the reddened spectra of HD\,183143 and 4U\,1907+09. 

\end{itemize}

Thus, a total of 14 NIR DIBs are confirmed at wavelengths longer than 9500~\AA; 
and up to 15 new NIR DIB candidates are reported -- 11 of which are discussed in this work for the first time. 

\paragraph{Properties and behaviour of NIR DIBs}

\begin{itemize}

\item Most of the NIR DIBs are intrinsically fairly narrow with FWHM $\sim$2--4\AA; only the 15610 and 17800\AA\ NIR DIBs are broader.

\item Although the NIR DIBs are fairly strong, their carrier abundance is about an order of magnitude smaller than the strongest optical DIBs.

\item The 11797, 13175, and 15268~\AA\ NIR DIBs are found to correlate well with the 5780~\AA~DIB, and by extension correlate 
better with \ion{H}{i} than with \Ebv. The strength ratio of 2 between the 5780 and 15268 DIBs implies that 
(N$f$)$_\mathrm{5780}$ / (N$f$)$_\mathrm{15268}$ = 14; however, the weak 15268 DIB toward HD~147889 precludes a common carrier 
for the two DIBs.

\item Laboratory spectra of large PAH cations show some coincidences with the NIR DIBs; however, their bands are probably to
  broad to explain the NIR DIBs.

\item The NIR DIBs in the translucent cloud towards HD\,147889 are nearly absent while most optical DIBs have average strength. 
  Since the ionisation fraction of small to medium-sized PAHs is low in this cloud, this hints at ionised species as the 
  carriers of the NIR DIBs.

\end{itemize}

The NIR range has proven a fertile wavelength range to search for new DIBs. 
Given our survey, most strong NIR DIBs between 10000~\AA\ and 25000~\AA\ are likely detected. 
However, numerous more NIR DIBs could be hiding in the noise, in particular in those regions affected most severely by residuals 
from telluric line removal procedures. Undoubtedly, further advances in technology, instrumentation and modelling of telluric 
and stellar atmospheres will lead to the detection of more NIR DIBs.

The NIR DIBs allows one to study the DIBs in highly extincted lines of sight that are inaccessible to optical spectroscopy. 
This offers some prospects to observe NIR DIBs and PAH infrared emission bands in the same environments. 
This would most certainly provide a critical test for the PAH-DIB hypothesis. 
To connect such studies back to optical DIBs, extensive studies are required of both optical and NIR DIBs in a large sample 
of sightlines, including detailed analyses of their relation to known molecular interstellar species as well.

\begin{acknowledgements}

We thank the referee for a thorough reading of the manuscript and several insightful comments
that improved the paper. NLJC thanks the staff at Paranal for their help in optimising the
execution of our programme, and X-shooter instrument scientists at ESO for assistance in data
handling issues.  JC acknowledges support from the Natural Sciences and Engineering Council
of Canada (NSERC). We thank Lucas Ellerbroek for guidance in using Spextool to correct the
data for telluric absorption lines. This research has made use of the Simbad database
operated by CDS in Strassbourgh.

\end{acknowledgements}

\bibliographystyle{aa}
\bibliography{/lhome/nick/Desktop/ReadingMaterial/Astronomy/Bibtex/bibtex} 

\end{document}